\documentclass[letterpaper,english,reprint]{revtex4-2}
\usepackage{lmodern}

\usepackage[T1]{fontenc}
\usepackage[latin9]{inputenc}
\usepackage{color}
\usepackage{babel}
\usepackage{amsmath}
\usepackage{amssymb}
\usepackage{graphicx}
\usepackage[unicode=true,pdfusetitle,
 bookmarks=true,bookmarksnumbered=false,bookmarksopen=false,
 breaklinks=false,pdfborder={0 0 1},backref=false,colorlinks=true]
 {hyperref}
 \usepackage[all]{hypcap} 
\hypersetup{
 linkcolor=blue, urlcolor=blue, citecolor=blue}

\makeatletter

\newcommand*\LyXThinSpace{\,\hspace{0pt}}
\pdfpageheight\paperheight
\pdfpagewidth\paperwidth

\usepackage{hyperref}
\usepackage[all]{hypcap}

\makeatother

\begin{document}
\title{Local and collective transitions in sparsely-interacting ecological
communities}
\author{Stav Marcus, Ari M. Turner and Guy Bunin}
\affiliation{Department of Physics, Technion-Israel Institute of Technology, Haifa
32000, Israel}
\begin{abstract}
Interactions in natural communities can be highly heterogeneous, with
any given species interacting appreciably with only some of the others,
a situation commonly represented by sparse interaction networks. We
study the consequences of sparse competitive interactions, in a theoretical
model of a community assembled from a species pool. We find that communities
can be in a number of different regimes, depending on the interaction
strength. When interactions are strong, the network of coexisting
species breaks up into small subgraphs, while for weaker interactions
these graphs are larger and more complex, eventually encompassing
all species. This process is driven by emergence of new allowed subgraphs
as interaction strength decreases, leading to sharp changes in diversity
and other community properties, and at weaker interactions to two
distinct collective transitions: a percolation transition, and a transition
between having a unique equilibrium and having multiple alternative
equilibria. Understanding community structure is thus made up of two
parts: first, finding which subgraphs are allowed at a given interaction
strength, and secondly, a discrete problem of matching these structures
over the entire community. In a shift from the focus of many previous
theories, these different regimes can be traversed by modifying the
interaction strength alone, without need for heterogeneity in either
interaction strengths or the number of competitors per species.
\end{abstract}
\maketitle
Interactions between species play important roles in shaping ecological
communities. A central challenge in community ecology is to relate
properties of interactions, such as their strength and organization,
to characteristics of communities such as diversity and response to
perturbations. In modeling, theory and simulations, some of the potential
interactions are assumed to be negligible or irrelevant and are taken
to be zero, a property known as sparseness.

Broadly speaking, theoretical approaches vary with the level of sparseness.
On the sparse side of this continuum, i.e., when many of the interactions
are zero, studying the structure of the network of interactions has
been fruitful \citep{guimaraes_Structure_2020}. Many phenomena have
been studied, including extended properties such as percolation, and
more local properties, such as the distribution of degree (number
of species interacting with each species). An extensive body of work
looks at local patterns within the network \citep{guimaraes_Structure_2020,bascompte_Disentangling_2009,holt_Indirect_2001,stone_Network_2019,milo_Network_2002}
known as network modules or motifs. Central and on-going questions
within this line of investigation include: whether these local patterns
are more common than some null expectation; whether they play a functional
role \citep{valverde_architecture_2018,holt_Theoretical_1997}; whether
it is possible to build-up from local properties to ecosystem-level
properties such as diversity \citep{fried_Communities_2016,bascompte_assembly_2009};
and whether the ignored ``weak'' links can indeed be neglected \citep{mccann_Weak_1998}.

In the other limit, when many or all possible interactions are present,
techniques have been developed \citep{may_Will_1972,bunin_Ecological_2017,kessler_Generalized_2015,fried_Communities_2016,allesina_Stability_2012,opper_Phase_1992,dougoud_feasibility_2018,tikhonov_Collective_2017,fisher_transition_2014,barbier_Generic_2018}
that relate the interaction strengths to properties such as the diversity,
existence of multiple stable states, and persistent dynamics. Here
two approaches have been used to model the community. In one, the
dynamics is linearized around a fixed point, and the parameters describing
the dynamics of coexisting species are sampled at random. This approach
predicts stability bounds \citep{may_Will_1972,allesina_Stability_2012},
and has been applied to sparse interactions \citep{mambuca_Dynamical_2021}.

In the other approach, known as community assembly, the dynamics of
species from a regional species pool is run, possibly resulting in
the extinctions of some of the species. One interesting observation
within the assembly approach, is that there are sharp transitions
in many-species communities, where persistent fluctuations, very many
alternative equilibria, or other properties emerge abruptly as relevant
interaction characteristics are changed \citep{opper_Phase_1992,may_Will_1972,fisher_transition_2014,kessler_Generalized_2015,fried_Communities_2016,bunin_Ecological_2017,tikhonov_Collective_2017}.
These characteristics are typically extended over the entire community
(e.g., moments of interaction strengths distribution) \citep{barbier_Fingerprints_2021}.
These transitions are known as collective transitions, because they
arise from community-wide processes, and a result of this is that
they become sharp in the many-species limit. Whether and how these
phenomena are found when interactions are very sparse (with a finite
number of links per species), and whether they are at all related
to local connectivity patterns that have been discussed for sparse
systems, has received little attention.

\begin{figure}
\begin{centering}
\includegraphics[width=1\columnwidth]{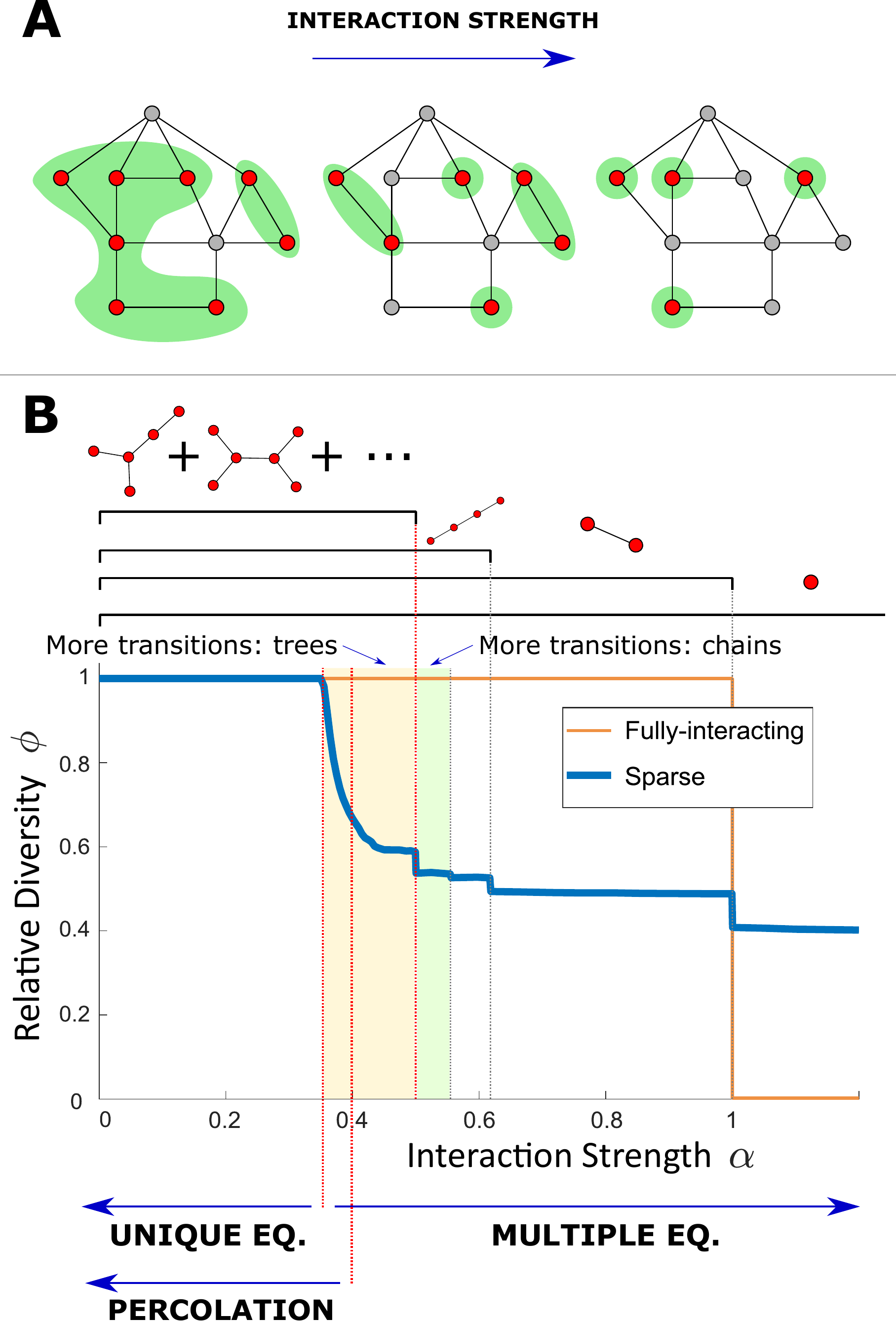}
\par\end{centering}
\caption{\textbf{\label{fig:phi(alpha)}Transitions in community structure.}
(A) Equilibria reached at different interaction strengths. Red vertices
represent persistent species, gray extinct species, and edges mark
pairs of interacting species. The persistent species can be divided
into connected subgraphs, shown with green background, separated by
the extinct species. As the interaction strength $\alpha$ is increased,
there are fewer and typically smaller allowed subgraphs, reducing
the number of coexisting species. (B) The relative diversity $\phi=S^{*}/S$
(where $S^{*}$ is the number of persistent species) at equilibrium,
from simulations with pool size $S=400$ and sparse interactions with
degree $C=3$ (blue); and for comparison for a fully-interacting community
(orange), which exhibits only a single jump at $\alpha=1$. The sparse
case exhibits infinitely-many sharp transitions, some of which are
marked by dashed vertical lines. By order of increasing $\alpha$,
the first transition is at $\alpha_{\mathrm{UE}}=\frac{1}{2\sqrt{2}}$,
from a unique to multiple equilibria. Next there is a percolation
transition at $\alpha_{\mathrm{perc}}\approx0.4$, below which a finite
fraction of the persistent species belong to a single giant connected
component. The other transitions result from changes in the allowed
connected subgraphs specifically those that are trees (see text),
leading to jumps in $\phi$. Some of the allowed trees are shown above
the graph. At values above $\alpha_{\mathrm{chain}}^{\left(\infty\right)}=\frac{1}{2}$,
the only allowed trees are chains of even length, each with a its
own allowed range. The allowed regions for trees that are not chains
terminate at different values of $\alpha$ depending on the tree,
all with $\alpha<1/2$ (shaded in orange).}
\end{figure}

Here we find that sparsely-interacting communities can exhibit phenomena
associated with both lines of investigation, in different regimes,
depending on interaction strength. In a theoretical model where a
community is assembled from a species pool, we study equilbria and
find that when interactions are strong, subgraphs of finite size play
a defining role in coexistence: the problem of species coexistence
reduces to a discrete problem on graphs involving local rules, in
the spirit of network motifs. The coexisting species can be separated
into connected subgraphs of the interaction network see Fig. \ref{fig:phi(alpha)}(A).
These play a central role in our theory. The number of possible subgraph
structures grows as the interaction strength is lowered, with the
subgraphs typically increasing in size. The addition of each new allowed
structure is marked by a transition in diversity and species abundance
distributions. Note that this would not be possible in a fully-interacting
community, which cannot break into multiple connected subgraphs.

At lower interaction strengths, as the interaction strength is varied
we find a percolation transition, and a transition between unique
and multiple alternative equilibria, similar to ones found in fully-interacting
systems \citep{biroli_Marginally_2018,diederich_Replicators_1989}.

Interestingly, all these phenomena do not require heterogeneity in
either the degree or the strength of interactions. In fact, the interactions
may even be locally ordered, that is, almost all species can have
identical neighborhoods up to a finite distance in the network. This
is in contrast to collective transitions studied previously, in which
heterogeneity is necessary for the transitions to occur \citep{may_Will_1972,bunin_Ecological_2017,kessler_Generalized_2015,fried_Communities_2016,allesina_Stability_2012,opper_Phase_1992,dougoud_feasibility_2018,tikhonov_Collective_2017}.
The interaction strength thus becomes an important parameter on its
own, divorced from the width of the distribution.

The paper is organized as follows. Sec. \ref{sec:constant sparse interactions}
introduces a theoretical model of a sparsely interacting competitive
community assembled from a species pool, in which each species interacts
with the same number of other species, and all interaction are of
identical strength. The properties of equilibria at different interaction
strengths are discussed. Interacting subgraphs of coexisting species
are introduced and their role is elucidated. Jumps in diversity, a
percolation and a unique- to multiple-equilibria transitions are found.
Sec. \ref{sec:Disorder-and-ER} extends the model to include heterogeneity
in the network of the vertex degree and interaction strengths. Sec.
\ref{sec:Hierarchy} shows how connected subgraphs form by combinations
of smaller ones. Sec. \ref{sec:Discussion} concludes with a discussion.

\section{Constant sparse interactions\label{sec:constant sparse interactions}}

\subsection{The model\label{sec:the model}}

We work within the framework of species assembly, where species migrate
from a species pool, and interact inside a community. The abundances
change in time according to the standard multi-species Lotka-Volterra
equations. There are $S$ species in the pool. The abundance of the
$i$-th species, $N_{i}$, follows the equation
\begin{equation}
\frac{dN_{i}}{dt}=r_{i}N_{i}\left(1-\sum_{j}\alpha_{ij}N_{j}\right)+\lambda_{i}.\label{eq:LV}
\end{equation}
where $\alpha_{ij}$ are the interaction coefficients, $r_{i}$ the
growth rates, and $\lambda_{i}$ the migration rates.

In this paper the matrix $\alpha_{ij}$, called the interaction or
community matrix \citep{levins_Evolution_1974,may_Stability_1973},
is always assumed to be symmetric, $\alpha_{ij}=\alpha_{ji}$, with
equal intraspecific competition for all species, $\alpha_{ii}=1$.
The symmetry ensures that the dynamics in Eq. (\ref{eq:LV}) always
reaches an equilibrium \citep{macarthur_Species_1970}; There may
be one or more such equilibria. Here we only consider competitive
interactions, $\alpha_{ij}\ge0$, and assume that all growth rates
are positive, $r_{i}>0$; other than that the values of the $r_{i}$'s
have no effect on the set of stable equilibria. In simulations we
take all $r_{i}=1$, and run Eq. (\ref{eq:LV}) until changes in the
$N_{i}$-values are small. The migration strengths $\lambda_{i}$
are taken to be small, $\lambda_{i}\rightarrow0^{+}$, ensuring that
at an equilibrium (i.e. a stable fixed point), all species that could
invade do so. We use a migration rate of $\lambda_{i}=10^{-10}$,
and species are considered extinct when $N_{i}<10^{-5}$. To ensure
a true equilibrium has been reached, it is verified that $1-\sum_{j}\alpha_{ij}N_{j}=0$
for all present species with $N_{i}>0$, and that extinct species
cannot invade, $dN_{i}/dt<0$.

We are interested here in sparse interactions, where many of the pairs
of species do not interact ($\alpha_{ij}=0$). The network of interactions
forms an undirected graph, with vertices representing species and
edges representing pairs of interacting species, sometimes called
the community graph \citep{logofet_Matrices_1993}.

It is common to use random interactions sampled from different distributions,
which capture different interaction characteristics. In this section
we will consider the following model: (1) Each species interacts with
exactly $C$ other species, with the interacting pairs chosen at random
so that the community graph is a random $C$-regular undirected graph.
(2) The interaction strength is equal for all interacting pairs. Therefore,
the interaction matrix can be written as $\alpha_{ij}=\delta_{ij}+\alpha A_{ij}$,
where $A_{ij}$ is the symmetric adjacency matrix of the community
graph. We consider $C\ll S$, and more precisely the limit of large
$S$ at constant $C$. We will see that this simplified model already
yields dramatically different results as compared with the fully-connected
system. Extensions to varying interaction strength and number of interaction
per species are then discussed in Section \ref{sec:Disorder-and-ER}.

We limit the discussion to properties of the system's equilibria,
and not the dynamics towards the equilibria, or under additional noise,
which are very interesting (some already discussed in \citep{bunin_Directionality_2021})
but beyond the scope of this work.

\subsection{\label{subsec:Overview-of-regimes}Overview of different regimes}

To get a bird's eye view of the different behaviors, we follow the
diversity at the equilibria as a function of the interaction strength
$\alpha$ (recall that in this first model $\alpha$ is identical
for all pairs). Let $S^{*}$ be the number of coexisting species at
an equilibrium (species richness), and define the relative diversity
$\phi=S^{*}/S$, their fraction relative to the total number $S$
of species in the pool. Fig. \ref{fig:phi(alpha)} shows simulation
results for $\phi$ as a function of $\alpha$. $\phi$ is estimated
by running simulations of Eq. (\ref{eq:LV}) over many realizations
of adjacency matrices $A_{ij}$, starting from a few different initial
conditions per realization, with each $N_{i}$ sampled uniformly from
$\left[0,1\right]$. The variability in $\phi$ between simulations
under the same conditions decreases with the diversity $S$, and for
large $S$ it is essentially set deterministically. For comparison,
the case of a full community matrix, where all species interact with
each other with strength $\alpha$ is also plotted. In this case the
behavior is simple: For $\alpha<1$ there is a unique fixed point
in which all species are persistent and $\phi=1$, while for $\alpha>1$
there are $S$ different fixed points, each with a single persistent
species so that $\phi=1/S$, tending to zero at large $S$. The sparsely-interacting
system, in contrast, is very rich and exhibits multiple different
behaviors with sharp transitions between them. At values of $\alpha$
close to zero, the interactions are weak enough to allow all species
to coexist with $\phi=1$. This persists for larger $\alpha$ up to
some critical value $\alpha_{\mathrm{UE}}$ where $\phi$ starts to
decrease. At another value $\alpha_{\mathrm{perc}}$ there is a percolation
transition, above which none of the components of persistent species
scales with the system size. The relative diversity $\phi$ keeps
decreasing until it reaches another transition where there is a jump
in $\phi$, at a value we denote by $\alpha_{\mathrm{chain}}^{\left(\infty\right)}$.
At $\alpha>\alpha_{\mathrm{chain}}^{\left(\infty\right)}$, the relative
diversity $\phi\left(\alpha\right)$ consists of infinitely many plateaus
punctuated by jumps, until the last jump at $\alpha=1$ and a single
plateau above it.

In the following sections, we discuss this behavior in detail, and
explain the multiple changes in system behavior and the reasons behind
them. We will show in the next sections that $\alpha_{\mathrm{UE}}=\frac{1}{2\sqrt{C-1}}$
and $\alpha_{\mathrm{chain}}^{\left(\infty\right)}=\frac{1}{2}$,
and provide analytical values for $\alpha$ of all jumps in $\phi\left(\alpha\right)$
at $\alpha\ge1/2$. In Subsection \ref{subsec:Percolation-phase-transition}
we discuss the percolation transition, and in Subsection \ref{subsec:UFP-to-MFP transition}
the unique to multiple equilibria transition, and show that it coincides
with $\alpha$ where $\phi$ first drops below 1.

\subsection{Allowed subgraphs and their dependence on interaction strength\label{subsec:Changes-in-allowed subgraphs}}

Here we begin to explain the different regimes described in Section
\ref{subsec:Overview-of-regimes}, by analyzing properties of the
equilibria of the model. In the limit of small migration ($\lambda_{i}\rightarrow0^{+}$)
some of the species will persist ($N_{i}>0$) and others go locally
extinct ($N_{i}=0$ as $\lambda_{i}\rightarrow0^{+}$). At an equilibrium,
the extinct species must be unable to invade ($dN_{i}/dt<0$), and
the abundances of the persistent species must return to the fixed
point if perturbed away from it. These conditions will be referred
to as \emph{uninvadability} and \emph{stability}, respectively. The
persistent species can be grouped into connected subgraphs of the
community graph, see Fig. \ref{fig:phi(alpha)}(A).

We begin in the limit of very large $\alpha$, studied in \citep{fried_Communities_2016,fried_Alternative_2017}.
Under this very strong competition, the problem reduces to two conditions.
First, two interacting species cannot both persist (competitive exclusion).
The connected subgraphs are thus individual species, see Fig. \ref{fig:phi(alpha)}(A),
rightmost illustration. Second, an extinct species cannot invade if
and only if it interacts with one or more persistent species. Stability
is automatically satisfied, as it involves isolated persistent species.
Importantly, the values of $\alpha$ do not appear in these two conditions,
and so finding an equilibrium point reduces to a discrete, combinatorial
problem on the graph, of finding a maximally independent set \citep{fried_Communities_2016}.
In \citep{fried_Alternative_2017}, the authors used this insight
to calculate the diversity and number of equilibria on Erd\H{o}s\textendash Rényi
graphs (where the pairs of interacting species are chosen independently
with some probability).

At lower values of $\alpha$ the connected subgraphs are no longer
only isolated species, see Fig. \ref{fig:phi(alpha)}(A). These subgraphs
must satisfy certain ``internal properties'' in order for them to
appear at a given $\alpha.$ As long as all of the neighboring species
to the subgraph are extinct, the abundances at a fixed point of the
subgraph are determined entirely by interactions within it. These
abundances must be positive (a condition known as \emph{feasibility}),
and the fixed point must be stable. These conditions depend only on
$\alpha$. This allows us to understand much of the behavior by looking
at individual subgraphs: each subgraph $\mu$ will have a critical
value $\alpha_{c}^{\left(\mu\right)}$, above which it is either unstable
or not feasible, and can therefore only appear at an equilibrium of
a system in the ``allowed'' range $\alpha<\alpha_{c}^{\left(\mu\right)}$.
(This leaves out a possibility that a graph could switch back and
forth between being allowed or not, see Appendix \ref{sec:Critical chain alphas}.)
Thus the system is governed by discrete combinatorial conditions,
which determine the entire set of possible equilibria of a given system.

Here another important simplification enters. Sparse random graphs,
including random-regular graphs and Erd\H{o}s\textendash Rényi graphs
discussed in Sec. \ref{sec:Disorder-and-ER}, are locally tree-like,
meaning that they have only a finite number of short cycles even when
$S$ is large. For example, in a large random regular graph with $C=3$
the average number of triangles is $4/3$ \citep{bollobas_probabilistic_1980}.
Thus, most connected subgraphs of finite size in the network will
be trees, i.e., contain no cycles, and properties such as diversity
and species abundance distribution that are averages over the entire
community can be calculated by only considering trees, and specifically,
the critical values $\alpha_{c}^{\left(i\right)}$ need to be found
only for trees. Examples of connected subgraphs within a local tree
neighborhood are shown in Fig. \ref{fig:Equilibrium-in-neighborhood}
in Appendix \ref{sec:Critical chain alphas}.

The trees can be divided into chains and other trees. We calculated
$\alpha_{c}^{\left(\mu\right)}$ for chains analytically, see Appendix
\ref{sec:Critical chain alphas}. For a chain with $n$ species,

\begin{equation}
\alpha_{c}^{\left(\mu\right)}\equiv\alpha_{\mathrm{chain}}^{\left(n\right)}=\begin{cases}
\frac{1}{2\cos\left(\frac{\pi}{n+1}\right)} & n\text{ even}\\
\frac{1}{2} & n\text{ odd}
\end{cases}\label{eq: alpha chains}
\end{equation}
For chains of even length, $\alpha_{\mathrm{chain}}^{\left(n\right)}$
is a decreasing series that converges from above to $\alpha_{\mathrm{chain}}^{\left(\infty\right)}=\frac{1}{2}$,
which is also the critical value for all chains of odd length, $\alpha_{\mathrm{chain}}^{\left(n\right)}=\frac{1}{2}$.
All other trees have $\alpha_{c}^{\left(\mu\right)}\leq\frac{1}{2}$,
with the first ones appearing, coincidentally, exactly at $1/2$,
as we prove in Appendix \ref{sec:Critical chain alphas}. Therefore,
$\alpha_{c}^{\left(\mu\right)}>\frac{1}{2}$ only for chains of even
length, so only they can appear in communities at $\alpha>1/2$.

In addition to these ``intrinsic'' considerations about the stability
and feasibility of different connected components, uninvadability
must also be considered. This is more complex since it depends on
how the components fit together, and in principle this could lead
to additional jumps in $\phi$, but in the $\alpha>1/2$ region, such
jumps seem to be rare if they exist at all, and their size is so small
that we have not detected them in simulations. See details in Appendix
\ref{sec:Extinct-species}.

This means that for $\alpha>1/2$, in ranges of $\alpha$ between
the $\left\{ \alpha_{c}^{\left(\mu\right)}\right\} $, the same trees
will be allowed and so essentially the same set of equilibria will
exist (since uninvadability does not seem to be important except at
the transitions). As $\alpha$ is lowered below some $\alpha_{c}^{\left(\mu\right)}$,
a new tree abruptly appears, leading to many new possible configurations
and thus causing the diversity to jump. While the dynamical simulations
used to obtain $\phi\left(\alpha\right)$ do not necessarily reach
all equilibria with the same probability, they clearly show jumps
in $\phi$ at these values, with plateaus of approximately constant
values of $\phi$ in between. Fig. \ref{fig:phi(alpha)}(B) shows
the function $\phi\left(\alpha\right)$, marking some of the critical
$\alpha_{\mathrm{chain}}^{\left(n\right)}$ from Eq. (\ref{eq: alpha chains})
as dashed vertical lines, showing that the jumps in $\phi$ indeed
happen exactly at $\alpha_{\mathrm{chain}}^{\left(n\right)}$. This
also happens for trees that are not chains when $\alpha<1/2$, see
an example in Appendix \ref{sec:Critical chain alphas}.

As $\alpha$ is lowered, infinitely many subgraphs of more complex
structures become stable, so the values $\left\{ \alpha_{c}^{\left(\mu\right)}\right\} $
become more dense, and the jumps in $\phi\left(\alpha\right)$ smaller
(see Fig. \ref{fig:non-chain histogram}). This makes it harder to
observe them in numerics, but we expect that they exist in the entire
range down to $\alpha_{\mathrm{UE}}$, defined in the following. Once
trees appear there are many interesting types of transitions that
could happen. Just as at $1/2$ arbitrarily long chains appear, there
could be other points where there are qualitative changes in the properties
of trees; see the Discussion section for more discussion.

\begin{figure}
\begin{centering}
\includegraphics[width=1\columnwidth]{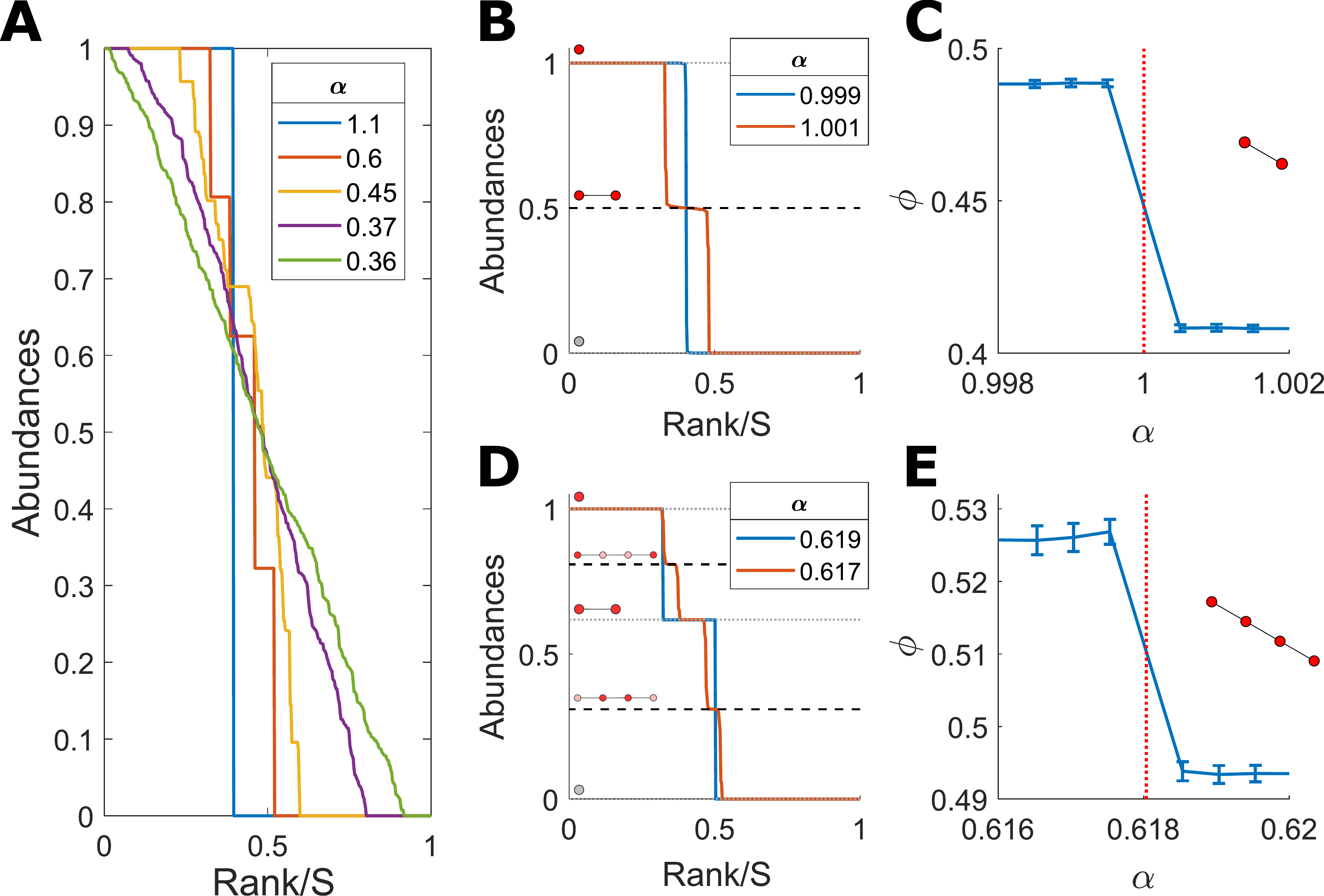}
\par\end{centering}
\caption{\label{fig:discontinuities zoom in}\textbf{Changes} \textbf{in feasible
and stable trees are reflected in species abundances. }At each value
of the interaction strength $\alpha$, certain trees are allowed,
and the abundance of a species depends only on $\alpha$ and the position
within a tree. \textbf{(A)} The rank-abundance curves at equilibria
reached dynamically for $S=400$,$C=3$, at several values of $\alpha$.
As $\alpha$ decreases, the increasing number of feasible and stable
trees generates more possible species abundances.\textbf{ (B-E)} Species
abundances on both sides of two transitions at $\alpha_{\mathrm{chain}}^{\left(n\right)}$
for $n=2,4$, where new trees appear. \textbf{(C)} and \textbf{(E)
}show the behavior of $\phi$ around the transitions associated with
pairs of species and chains of length 4 respectively becoming feasible
and stable.\textbf{ (B)} and (\textbf{D)} show the abundances at equilibria
at values of $\alpha$ on two sides of the transitions. The expected
abundances are marked by dashed black lines, with thicker lines for
the abundances of the species in the tree associated with the transition.
Next to each abundance appears the tree that contains it, with the
species that have this abundance in dark red (or gray in the case
of the abundance 0 of extinct species).}
\end{figure}

The transitions are also reflected in the possible abundances of species,
as seen in rank-abundance curves, which show the abundances sorted
in decreasing order, see Fig. \ref{fig:discontinuities zoom in}.
At a given $\alpha$, the abundance of a species depends only on the
connected tree it belongs to, and its position within it; for example,
species that belong to a chain of length two have $N_{i}=\frac{1}{1+\alpha}$.
Therefore, as a tree $\mu$ becomes feasible and stable at $\alpha=\alpha_{c}^{\left(\mu\right)}$,
the abundances associated with it can appear at an equilibrium. As
shown in Fig. \ref{fig:discontinuities zoom in}(A), this causes the
abundance graphs to smooth out as $\alpha$ is lowered, since the
number of possible abundances increases.

To summarize, in this section we described how the interaction network
breaks up into connected subgraphs, with changes in allowed subgraphs
driving jumps in diversity and species abundances. These subgraphs
are trees that are feasible and stable at that interaction strength.
Finding the equilibria of Eq. (\ref{eq:LV}) reduces into a discrete
graph theoretical problem on the community graph. Broadly speaking,
for stronger competition there are fewer and typically smaller allowed
trees.

As $\alpha$ is lowered, the size of the allowed subgraphs grows until
they span a finite fraction of the species, as discussed in the next
section. The number of different types of allowed graphs quickly grows
with their size, and the problem of classifying them becomes more
difficult, and less useful. These very large connected graphs can
include the rare but still existing cycles in the graphs, and so they
are no longer trees.

\subsection{Percolation transition\label{subsec:Percolation-phase-transition}}

Percolation transitions are one of the canonical phenomena studied
in graph theory. In site percolation, some vertices of a graph are
removed. As the probability of vertex removal varies, on one side
of the transition the remaining graph breaks into small (sub-extensive)
pieces; on the other side, a finite fraction of vertices belong to
a single connected component. Natural communities belonging to both
regimes are known to exist \citep{guimaraes_Structure_2020}.

\begin{figure}
\begin{centering}
\includegraphics[width=1\columnwidth]{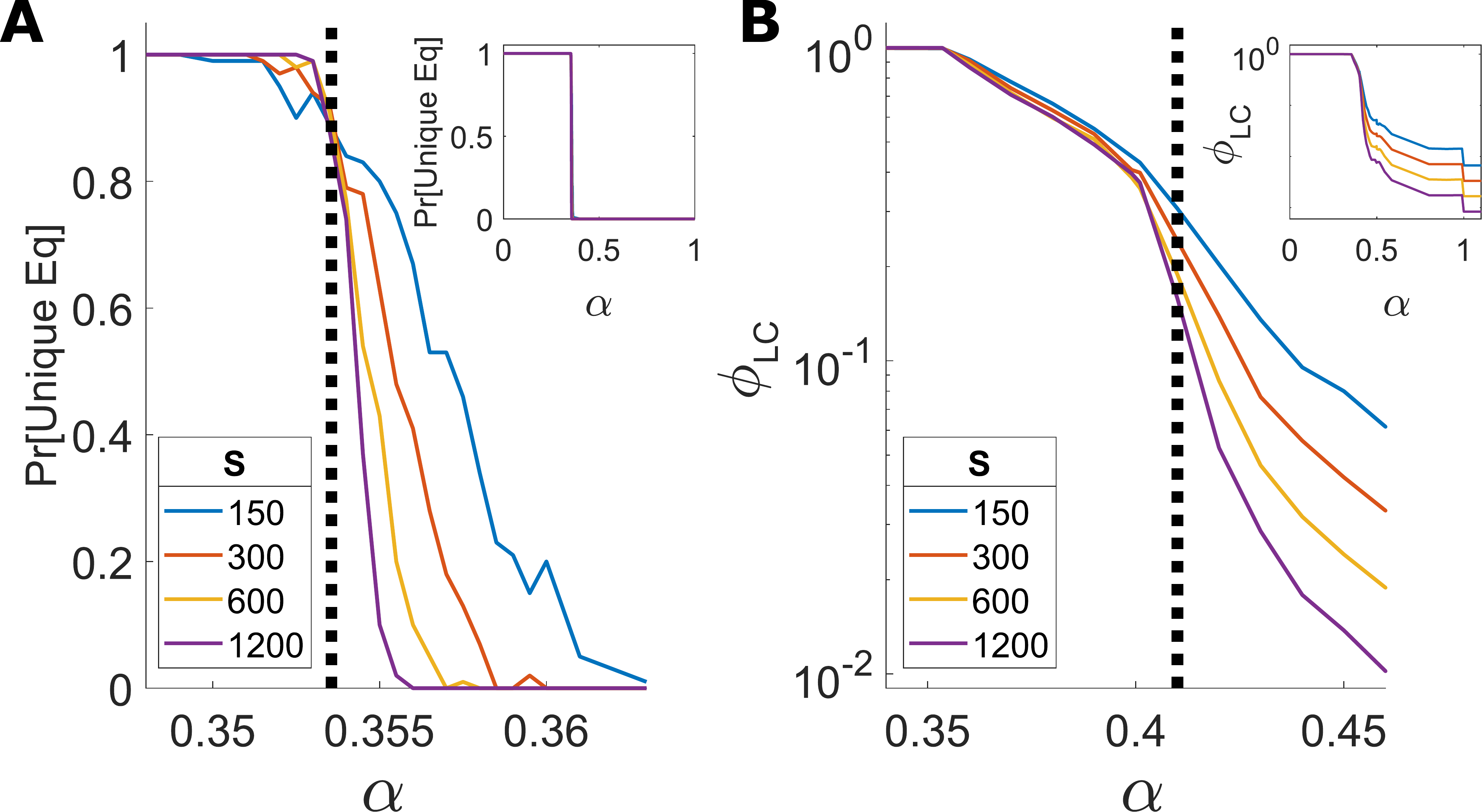}
\par\end{centering}
\caption{\label{fig:UFP-MFP and PERC phase transition}\textbf{Collective transitions.}
\textbf{(A)} Unique to multiple equilibria transition: the probability
for a unique equilibrium as a function of $\alpha$, for connectivity
$C=3$ and several pool sizes $S$. The probability is obtained by
generating many realizations of interaction matrices and determining
whether there is a unique equilibrium by the stability of the fully-feasible
fixed point, as described in the text body. The exact value for the
transition is shown as a dashed black line, Eq. (\ref{eq: alpha UFP}).
Inset: the same graph over a larger range of $\alpha$. The transition
becomes sharper with $S$ grows, as expected from a colective transition.
\textbf{(B)} Percolation transition: the fraction of species in the
largest connected component as a function of $\alpha$, for several
values of $S$. The location of the transition is $\alpha_{\mathrm{perc}}\approx0.41\pm0.01$
(dashed black line). At $\alpha<\alpha_{\mathrm{perc}}$ a finite
fraction of species belongs to the largest component even when $S$
grows. At $\alpha>\alpha_{\mathrm{perc}}$, this fraction decreases
with $S$. Inset: the same graph over a larger range of $\alpha$.
Here too, the transition becomes sharper at larger values of $S$.}
\end{figure}

We find that at some interaction strength $\alpha_{\mathrm{perc}}$
there is a percolation transition, below which the largest connected
subgraph formed by surviving species is extensive, that is, includes
a finite fraction of all the species. Fig. \ref{fig:UFP-MFP and PERC phase transition}(B)
shows the fraction of species belonging to the largest connected component
as a function of $\alpha$, for several values $S$ with $C=3$. Above
a certain $\alpha$, which for this connectivity is at $\alpha_{\mathrm{perc}}\left(C=3\right)\approx0.41\pm0.01$
(marked by a dashed line), this fraction drops with $S$ indicating
a sub-extensive largest component. Below $\alpha_{\mathrm{perc}}$
this fraction converges to a constant value. As expected, this value
is smaller than $1/2$, since at $\alpha>1/2=\alpha_{\text{chain}}^{\left(\infty\right)}$
the only possible components are finite-length chains, as shown in
Sec. \ref{subsec:Changes-in-allowed subgraphs} above. Also, $\alpha_{\mathrm{perc}}\ge\alpha_{\mathrm{UE}}$
where all species persist, see Sec. \ref{subsec:UFP-to-MFP transition}
below. The fact that the transition becomes sharper with growing $S$
is a hallmark of a collective transition.

Fig. \ref{fig:UFP-MFP and PERC phase transition}(B) is qualitatively
similar to that of a standard site-percolation transition, where vertices
are randomly and independently chosen to be ``present'', see Appendix
\ref{sec:classical percolcation}. However, the fraction of persistent
species at $\alpha_{\mathrm{perc}}$ is around $\phi_{\mathrm{perc}}\approx0.64\pm0.02$,
which is larger than the $\phi_{\text{perc}}=1/2$ of a standard site
percolation transition at $C=3$ \citep{bunde_Fractals_1996}. This
is because in our model, the species that persist are not sampled
independently; the higher value in our model is expected given that
persistent species are correlated, tending not to be adjacent to one
another.

\subsection{Unique to multiple equilibria transition\label{subsec:UFP-to-MFP transition}}

The final transition in the model with all-equal $\alpha$, at the
lowest value of $\alpha$, is from multiple to unique equilibria.
In order to find the critical value of $\alpha$ for this transition,
we first argue that the community has a unique equilibrium exactly
when it is ``fully feasible'', i.e. all species are persistent ($\phi=1$);
if the fully-feasible state is an equilibrium then it is necessarily
unique. Thus, the transition from the multiple equilibria phases to
the unique equilibrium phase occurs at the value $\alpha_{\mathrm{UE}}\left(C\right)$
in which $\phi$ drops below 1. The equivalence holds only for this
model where all species have the same number of interacting pairs
and all interactions have the same strength $\alpha$, and breaks
in more general cases, see Sec. \ref{sec:Disorder-and-ER} below.

To understand this relation, consider the Lyapunov function $F=2\sum_{i}N_{i}-\sum_{ij}N_{i}\alpha_{ij}N_{i}$,
which for symmetric interactions ($\alpha_{ij}=\alpha_{ji}$) grows
with time according to the Lotka-Volterra equations \citep{macarthur_Species_1970},
and whose local maxima coincide with the equilibria. The fixed point
where all species persist is always feasible, as from the local homogeneity
of the community graph all abundances are equal $N_{i}=\frac{1}{1+C\alpha}>0$,
and this would be stable if the full interaction matrix $\alpha_{ij}$
is positive definite. As $\alpha_{ij}$ is also the matrix of second
derivatives of $F$, if the fixed point is stable then the Lyapunov
function is concave everywhere, meaning the minimum at the ``fully
feasible'' equilibrium is global and therefore unique.

Conversely, if the fully feasible equilibrium is not stable, then
$F$ is a non-concave quadratic function on the quadrant $\{\forall i:N_{i}\geq0\}$
and one expectsthat if there are many potential species, it is likely
to have many local maxima, and therefore multiple equilibriaWe checked
this relation numerically, by generating 100 realizations of the interaction
matrix at a given $\alpha$, solving the dynamics in Eq. (\ref{eq:LV})
with $30$ different randomly chosen initial conditions, and checking
whether all runs converge to the same equilibrium. This process was
repeated around the transition (whose position is given below in Eq.
(\ref{eq: alpha UFP})), for $\alpha\in\left[0.35,0.36\right]$ when
$C=3$ and for $\alpha\in\left[0.285,0.3\right]$ when $C=4$, and
with $S=200,400$. In all runs, there was a unique fixed equilibrium
at exactly the same realizations that were fully feasible.

The stability is thus determined by the range in which the matrix
$\alpha_{ij}=\delta_{ij}+\alpha A_{ij}$ is positive definite. $A_{ij}$
is an adjacency matrix of a $C$-regular graph of size $S$, and at
large $S$ its minimal eigenvalue is with probability one at $\lambda_{\mathrm{min}}^{A}=-2\sqrt{C-1}$
\citep{mckay_expected_1981}. The minimal eigenvalue of the matrix
$\alpha_{ij}$ is therefore at $\lambda_{\mathrm{min}}=1+\alpha\lambda_{\mathrm{min}}^{A}=1-2\alpha\sqrt{C-1}$,
and the critical value of $\alpha$ will be

\begin{equation}
\alpha_{\mathrm{UE}}\left(C\right)=\frac{1}{2\sqrt{C-1}}\ .\label{eq: alpha UFP}
\end{equation}

Fig. \ref{fig:UFP-MFP and PERC phase transition}(A) shows the probability
of the system having a unique equilibrium as a function of $\alpha$
for several values of $S$, using the stability of the matrix $\alpha_{ij}$.
As $S$ increases, the probability for a unique equilibrium becomes
sharper (again, a clear sign of a collective transition), approaching
a step function at the expected value of the transition $\alpha_{\mathrm{UE}}\left(C\right)$.

\section{Heterogeneity in vertex degree and interaction strength\label{sec:Disorder-and-ER}}

\begin{figure}
\begin{centering}
\includegraphics[clip,width=1\columnwidth]{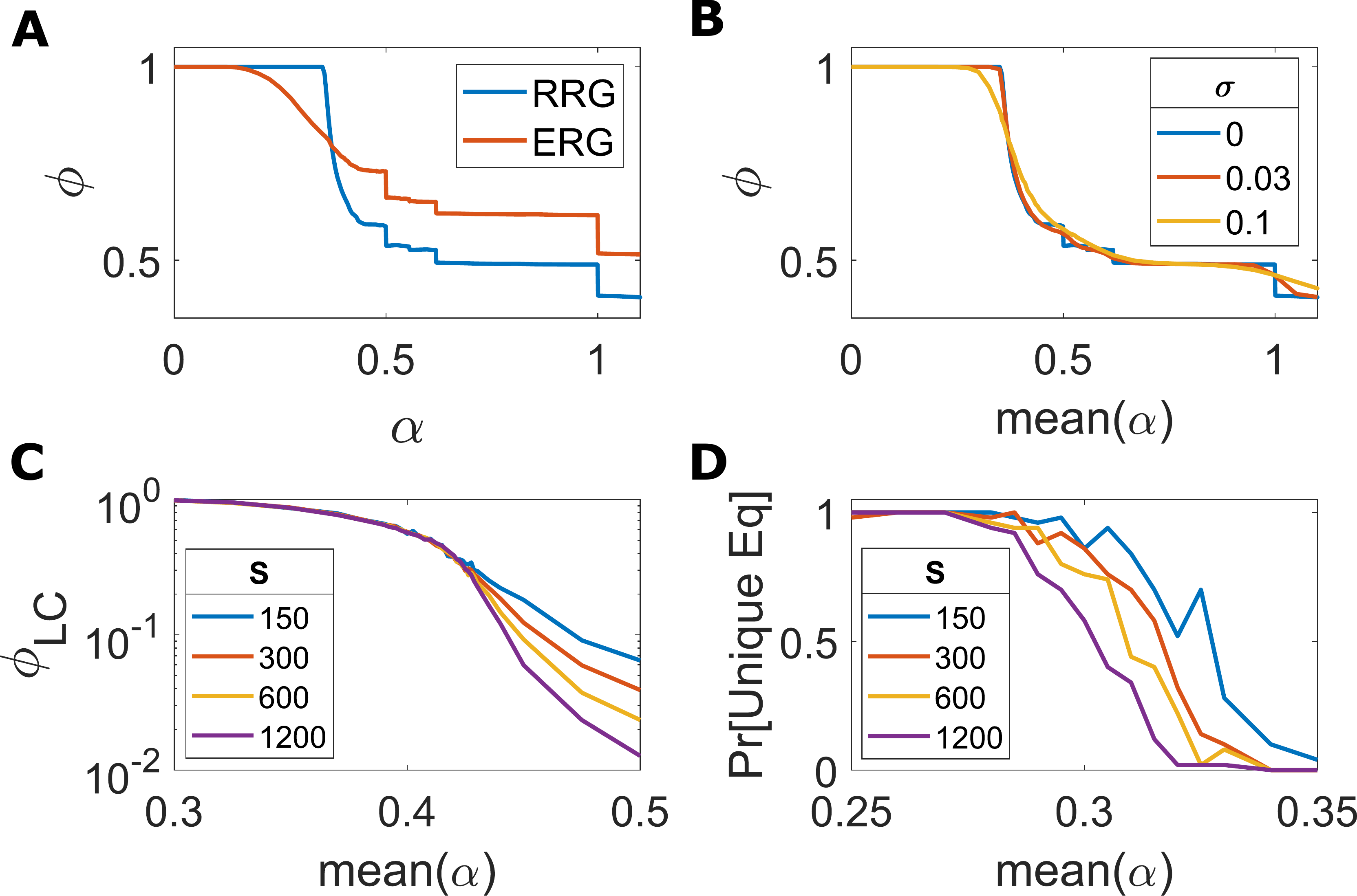}
\par\end{centering}
\caption{\label{fig:phi of alpha with noise}\textbf{Transitions with heterogeneous
interaction strengths and degrees. (A-B) }Transitions due to changes
in allowed trees are broadened when there is variability in interaction
strengths, but remain sharp for variation in degree.\textbf{ }The
relative diversity $\phi$ as a function of interaction strength $\alpha$,
for $S=400,C=3$.\textbf{ (A)} Erd\H{o}s-Rényi graphs with interaction
probability $p=C/S$, compared to a random $C$-regular graph. \textbf{(B)}
Interaction strength is drawn from a normal distribution with mean
$\alpha$ and standard-deviation $\sigma=0,0.03,0.1$, keeping the
interactions symmetric and a random regular graph. \textbf{(C-D})
The collective transitions with heterogeneity in interaction strength
become sharper as $S$ increases, just as they do without it (compare
with Fig. \ref{fig:UFP-MFP and PERC phase transition}). Results are
shown for\textbf{ $C=3$, $\sigma=0.1$ }and several values of $S$.
\textbf{(C) }Percolation transition: The fraction of species in the
largest connected component as a function of $\alpha$. \textbf{(D)}
Unique to multiple equilibria transition: The probability of having
a unique equilibrium as a function of $\alpha$.}
\end{figure}

So far, Sec. \ref{sec:constant sparse interactions} analyzed a model
where each species interacts with exactly $C$ others, and all with
the same interaction strength $\alpha$. Here we consider the effects
of heterogeneity, both in the strength of species interactions and
in the vertex degree (the number of species interacting with a given
one). The interaction strength is varied by drawing it from a normal
distributions with mean $\alpha$ and a given standard deviation $\sigma$.
The degree is varied by replacing the random regular graphs with an
Erd\H{o}s-Rényi random graph, in which each pair of species is independently
chosen to interact, such that the average degree is $C$. To understand
how these two changes affect the results, we consider them separately.
Fig. \ref{fig:phi of alpha with noise} shows the relative diversity
$\phi$ as a function of $\alpha$, for both cases.

When varying the degree, the jumps in the relative diversity $\phi$
due to changes in the allowed trees remain sharp, while they are broadened
for variations in interactions strength. This makes sense, as the
trees can still exist if the degrees vary; there may be additional
adjacent species but these do not affect whether the populations on
the tree are feasible and stable. On the other hand, the interaction
strengths affect the stability and feasibility of the tree. In an
Erd\H{o}s-Rényi graph, all trees of the same topology will all have
the same $\alpha_{c}$. ($\phi$ is different between the Erd\H{o}s-Rényi
and regular graphs due to their different structure.)

If interaction strengths are varied, trees in the same system, which
have the same topology but different interaction strengths might have
different limits on stability and feasibility, leading to the appearance
of more types of allowed trees than in the all-equal $\alpha$ case.
But if the disorder is not too strong, the picture of the all-equal
$\alpha$ case remains relevant: if the mean value of $\alpha$ is
within a plateau of the all-equal $\alpha$ case and not too close
to the ends, the interactions would mostly allow the same trees as
they would in the case without disorder. For example, for $\text{mean}\left(\alpha\right)=0.7$,
within the plateau allowing only pairs and singlets, for $\sigma=0.1$
these make up $99.5\%$ of feasible trees in a typical equilibrium
for large $S$. Indeed, for $\sigma=0.1$, in most of the range within
this plateau, $\phi$ is almost identical to the all-equal $\alpha$
case Fig. \ref{fig:phi of alpha with noise}(B).

The two remaining transitions, for percolation and from unique to
multiple equilibria, both appear to become sharper as $S$ increases
for both variations in interaction strengths and degree, as can be
seen in Fig. \ref{fig:phi of alpha with noise}(C-D) and in Appendix
\ref{sec:C Global transitions with heterogeneity}, Fig. \ref{fig:percolation and UFP/MFP with noise}.
For any given $S$ the transitions are broader compared to the equal-$\alpha$
model (Fig. \ref{fig:UFP-MFP and PERC phase transition}). Furthermore,
in both cases $\phi$ drops below 1 while still at the unique equilibrium
phase, which happens when the system is no longer fully feasible.
This is clearly seen for the Erd\H{o}s-Rényi random graph in Fig.
\ref{fig:phi of alpha with noise}(A), and is shown for varying interaction
strengths in Appendix \ref{sec:C Global transitions with heterogeneity}.
This is in contrast to the all-equal $\alpha$ model, where $\phi$
drops below 1 when the system becomes unstable at $\alpha_{\mathrm{UE}}$.

\section{Subgraph emergence rule: How the trees grow\label{sec:Hierarchy}}

As the interaction strength is lowered (by lowering $\alpha$ in Sec.
\ref{sec:the model} or $\text{mean}\left(\alpha\right)$ in Sec.
\ref{sec:Disorder-and-ER}), the allowed connected subgraphs become
larger (containing more species) and more complicated, until one connected
subgraph can take up a finite fraction of the community at the percolation
transition. Continuing to grow beyond that, they finally include the
entire network. For $\alpha>1/2$ there is a clear regularity in the
sequence of transitions, as even-length chains become allowed by order
of length. This raises the question of whether there is any regularity
by which more complicated subgraphs (trees, and even subgraphs with
cycles) become allowed. We now describe a general and simple result,
when the interactions strengths are heterogeneous.

\begin{figure}
\begin{centering}
\includegraphics[width=0.8\columnwidth]{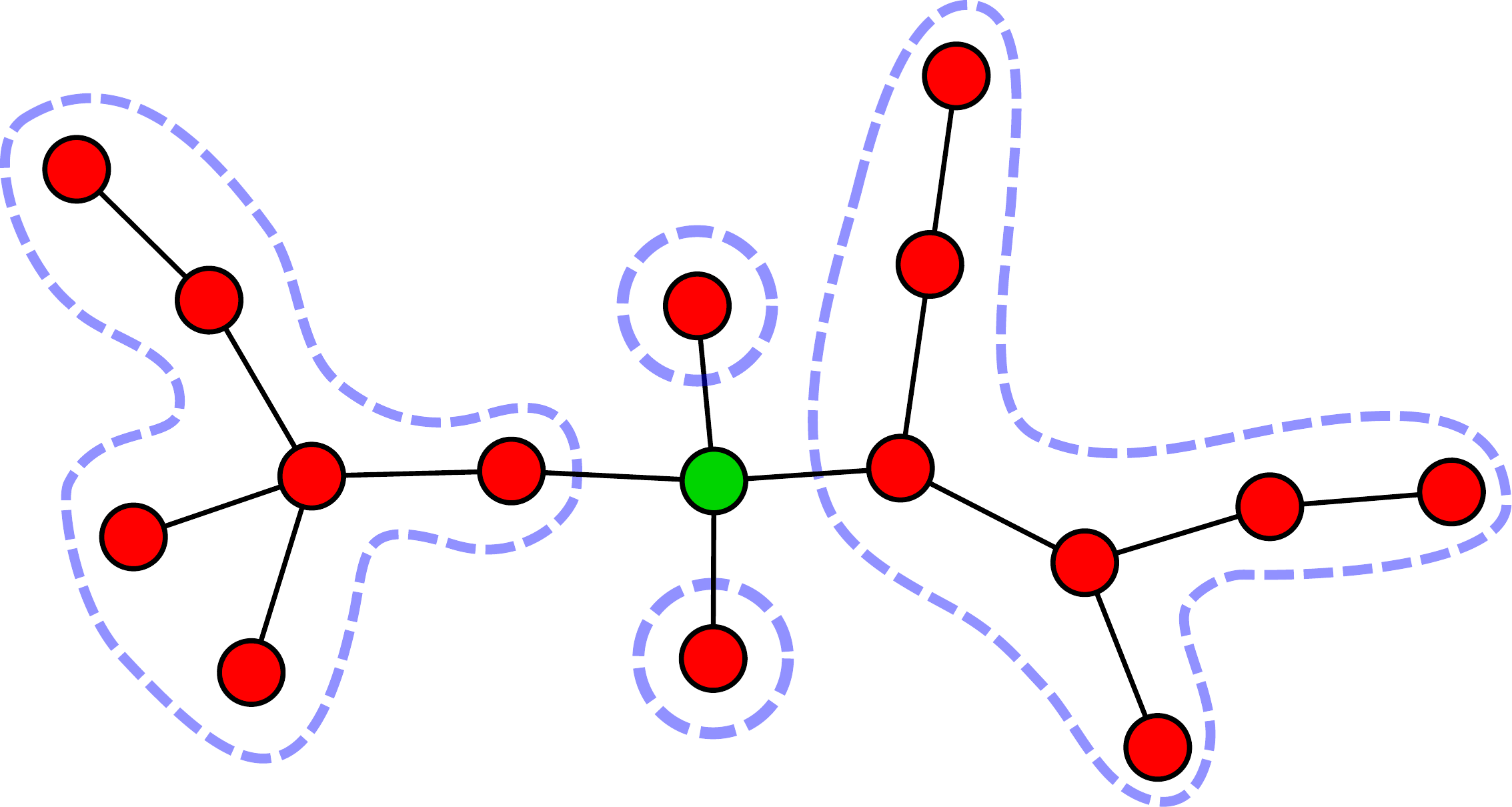}
\par\end{centering}
\caption{\label{fig:Hierarchy}A tree that becomes feasible and stable at some
$\alpha_{c}$, can be constructed from three or more trees that are
allowed right above $\alpha_{c}$ (surrounded by dashed lines), joined
by one additional species (green). This is true with probability one
when there is heterogeneity in interaction strengths.}
\end{figure}

Consider a subgraph within the interaction network, see Fig. \ref{fig:Hierarchy}.
Since interaction strengths are not all equal, this refers to a specific
set of vertices, which means the result can be different for the same
subgraph structure when it involves different species. To define $\alpha_{c}$
of the subgraph, consider the process by which the mean strength is
changed by shifting the values of the $\alpha_{ij}$, i.e. adding
a constant (other continuous changes of the matrix $\alpha$ are also
possible). Just below $\alpha_{c}$ all abundances are positive. We
prove in Appendix \ref{sec:App-Hierarchy} that with probability one,
it is feasibility, rather than stability, that is lost at $\alpha_{c}$,
by one species going extinct ($N_{i}\rightarrow0$). When this becomes
extinct, the remaining subgraph is composed of allowed subgraphs,
hence the subgraph right below $\alpha_{c}$ is thus composed of subgraphs
allowed right above $\alpha_{c}$, with one additional vertex.

For a tree subgraph, the species that goes extinct interacts with
three or more species in that subgraph (is a branching point), assuming
that the distribution of the $\alpha_{ij}$-values is not too wide.
See argument in Appendix \ref{sec:App-Hierarchy}. This implies that
the tree splits into three or more trees.

This construction gives a constraint on what order the specific subgraphs
become allowed, i.e. become feasible and stable: that when a subgraph
becomes allowed as $\alpha$ is decreased, pieces of it with one species
removed were already allowed. As noted above, because of the heterogeneity
of the interactions, here a subgraph refers to a specific set of species
on which it resides. Note that as always, whether an allowed graph
appears in an equilibrium depends also on the neighboring species
and the rest of the network.

\section{Discussion\label{sec:Discussion}}

We have looked at a community assembled from a pool of sparsely-interacting
species. When the interactions are strong enough, the assembly process
breaks the network into many connected subgraphs. The problem of equilibrium
coexistence reduces to understanding which subgraphs are allowed,
and how they are organized to keep extinct species from invading.

When these subgraphs are small, it might be possible to formulate
predictive local rules about their occurrence, in the spirit of ``assembly
rules'' \citep{stone_Network_2019,weiher_Ecological_2001,friedman_Community_2017}.
The simplest example is competitive exclusion, where if the interactions
between two species are greater than one, $\alpha_{ij},\alpha_{ji}>1$,
then they cannot coexist within a community of species that interact
competitively, irrespective of the state of the other species. This
can be interpreted as a rule that when interaction are stronger than
one, the connected components include just one species. Here this
regularity extends to weaker interaction strengths, first identifying
a regime where interacting pairs are also allowed, which is quite
robust to some level of heterogeneity in interactions strengths (Sec.
\ref{sec:Disorder-and-ER}), and then to regimes with larger connected
subgraphs.

For lower interaction strengths there are many larger allowed subgraphs,
making the corresponding graph-theoretical problem hard and far less
local, and limiting the potential for predictive local rules. There
is also more sensitivity to heterogeneity in the interaction strengths.
At even lower interaction strengths, connections percolate across
the entire network of coexisting species, and below that there is
a dramatic transition in behavior, as the equilibrium becomes unique,
similar to transitions found in fully-interacting systems \citep{diederich_Replicators_1989,bunin_Ecological_2017,biroli_Marginally_2018}.
We have not observed any sharp changes occurring at the percolation
transition, to diversity, stability or other measures beyond the graph
connectivity; percolation might however be a necessary bridge between
the finite-subgraph regime, and the unique to multiple-equilibria
transition.

In a striking difference from fully-connected networks, this rich
phenomenology does not require heterogeneity in interaction strengths
or vertex degrees, which are necessary in fully-interacting networks,
and have been central to much of the field for decades \citep{may_Will_1972,bunin_Ecological_2017,kessler_Generalized_2015,fried_Communities_2016,allesina_Stability_2012,opper_Phase_1992,dougoud_feasibility_2018,tikhonov_Collective_2017}.
This makes the interaction strength (or its mean as opposed to the
width of the distribution) a parameter of independent importance,
single-handedly driving changes in stability and feasibility. When
heterogeneity is present, the mean and distribution of interaction
strengths have a combined effect, with the allowed subgraphs still
playing a central role in shaping the community.

There are many mathematical questions to explore in these systems,
which are interesting because of the interplay between the combinatorial
structure of the community graphs and the quantitative properties
of the interaction matrices. Such questions include a further understanding
of the sequence of transitions: Are there other limit points where
infinite trees become stable (such as the infinite chain becoming
stable at $\alpha=1/2$)? And are there ranges where the critical
points $\alpha_{c}^{\left(\mu\right)}$ are dense? Another question
is how much the dynamics affects the distribution of equilibria reached,
as compared to an unbiased choice among all the allowed equilibria;
for example, how much this affects the sizes of the jumps in diversity.
Finally, it would be especially interesting to understand the transition
to the unique equilibrium state, by studying the structure of the
small groups of species that go extinct just above the transition.
It would likely be possible to make progress on many of these questions
by studying ideal infinite trees with a fixed degree $C$, so that
the inhomogeneity in the equilibria arises from their instabilities.

The extent to which interactions in different natural communities
are sparse is an open question, since directly measuring interaction
strengths can be hard, especially the weaker ones. This is complicated
by additional factors, as many weak interactions might have a large
cumulative effect, and that some inference techniques assume that
the network is sparse (e.g., \citep{kurtz_Sparse_2015}). One can
hope that studying consequences of sparsity would help identify and
better understand such communities.
\begin{acknowledgments}
It is a pleasure to thank M. Barbier, J. Friedman and J. Gore, for
stimulating discussions and helpful feedback. G.\LyXThinSpace Bunin
was supported by the Israel Science Foundation (ISF) Grant No. 773/18.
\end{acknowledgments}

\appendix

\section{\label{sec:Critical chain alphas}Critical values of trees}

\begin{figure*}
\begin{centering}
\includegraphics[width=0.8\paperwidth]{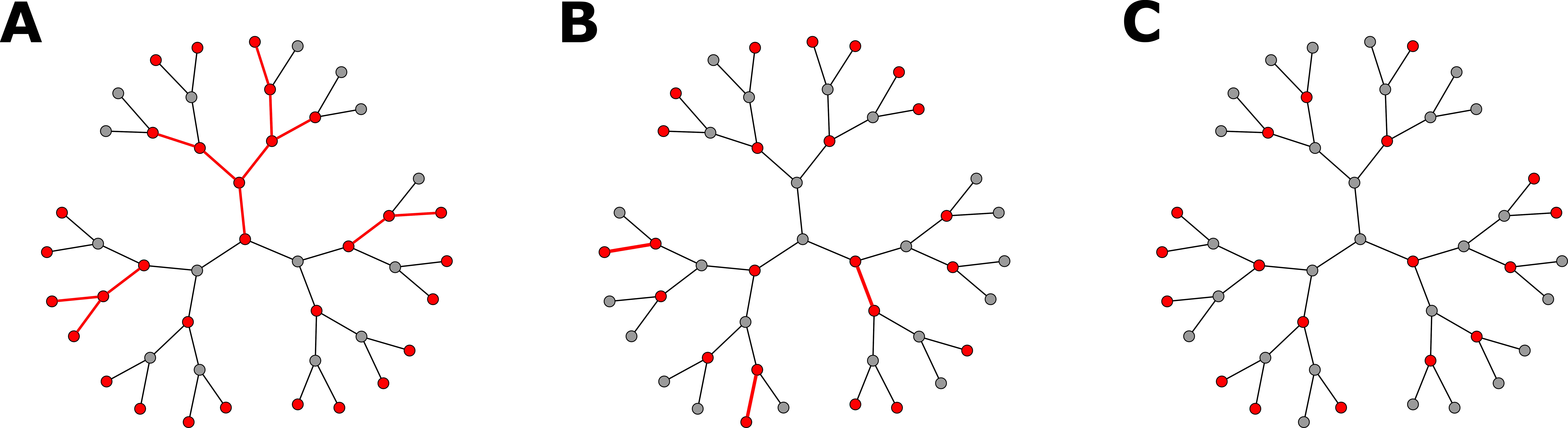}
\par\end{centering}
\caption{\textbf{\label{fig:Equilibrium-in-neighborhood}Equilibrium in the
local neighborhood of one species.} The neighborhood of one species
in a specific system with connectivity $C=3$ and pool size $S=1200$,
where the interaction strengths are \textbf{(A)} $\alpha=0.45$ \textbf{(B)}
$\alpha=0.7$ \textbf{(C)} $\alpha=1.1$. Extinct species are in gray,
and persistent species in red. The edges connecting two persistent
species are also marked in red. As the interaction strength is increased,
the subgraphs change from large trees that are not chains in (A),
to length-2 chains and singlets in (B), and singlets only in (C).}
\end{figure*}

The critical $\alpha$ values for connected trees, $\left\{ \alpha_{c}^{\left(\mu\right)}\right\} $,
are discussed in section \ref{subsec:Changes-in-allowed subgraphs}
of the main text. They are the values above which each tree becomes
either unstable or unfeasible, and therefore cannot appear in an equilibrium.
In rare cases, a stable tree can regain unfeasibility after losing
it (see below); here we define $\alpha_{c}^{(\mu)}$ more precisely
as the \emph{highest} value of $\alpha$ where the tree is feasible
and stable. As discussed in the main text, trees are important subgraphs
because the local neighborhoods of most species are locally tree-like.
Examples of trees appearing in an equilibrium in the neighborhood
of one species within a large community are shown in Fig. \ref{fig:Equilibrium-in-neighborhood}.

Stability changes at a single value of $\alpha$, so that the system
is stable for all values of $\alpha$ below it and unstable for all
values above it. Indeed, as the interaction parameters, in matrix
form, are represented by $I+\alpha A$, where $I$ is the identity
matrix (see Sec. \ref{sec:the model}), if the minimal eigenvalue
of the adjacency matrix $A$ representing a tree is $\lambda_{min}$,
the smallest eigenvalue of the tree at the interaction strength $\alpha$
would be $1+\alpha\lambda_{min}$, so the tree is unstable exactly
for $\alpha>-1/\lambda_{min}$. Feasibility on the other hand can
be gained and then lost more than once, but we find that most stable
trees that gain feasibility as $\alpha$ is lowered usually retain
it, with feasibility gained and lost again in only 0.03\% of trees
up to 20 vertices for $C=3$ and 0.01\% of trees up to 15 vertices
for $C=5$.

\begin{figure}
\begin{centering}
\includegraphics[width=1\columnwidth]{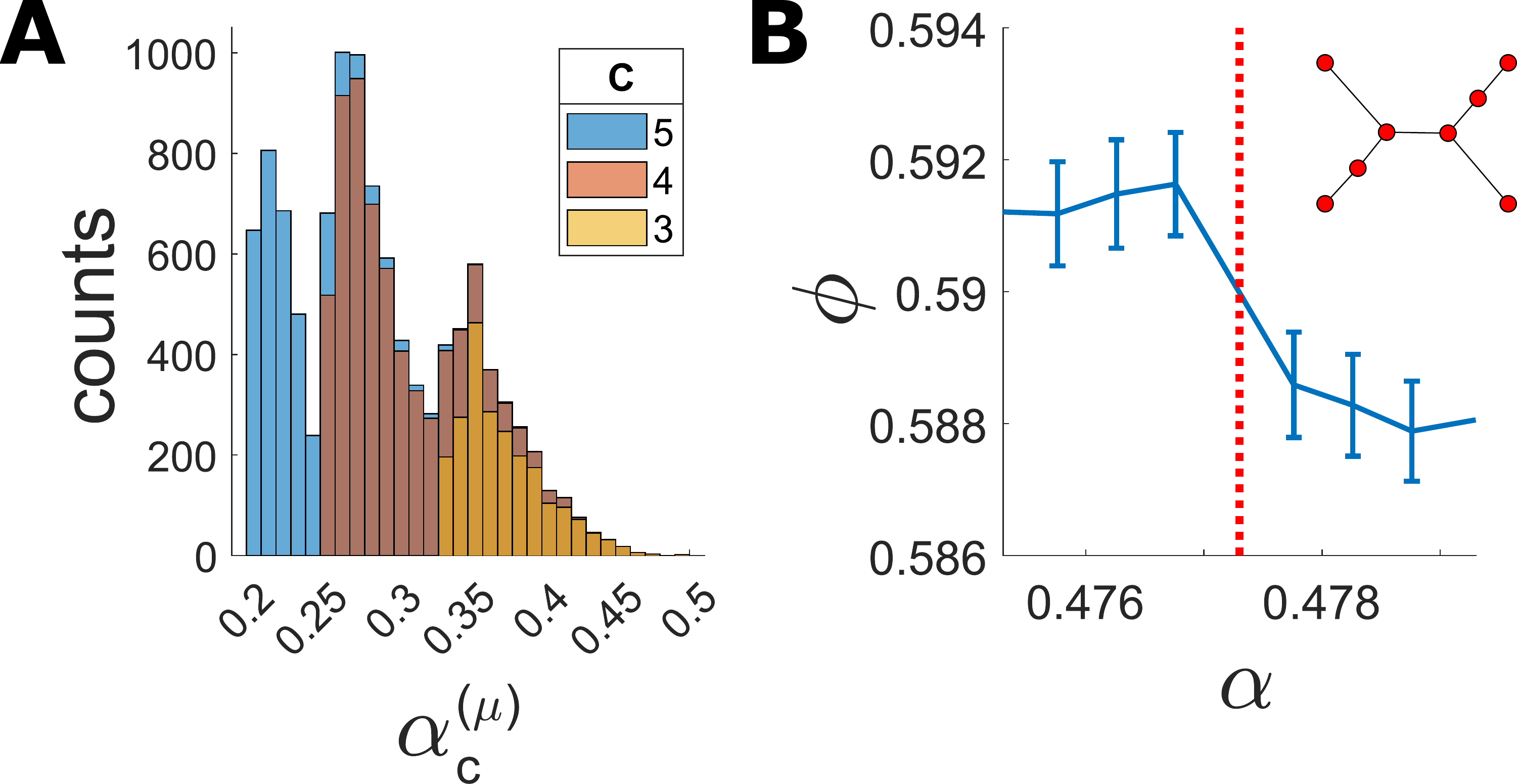}
\par\end{centering}
\caption{\label{fig:non-chain histogram} \textbf{The critical values $\left\{ \alpha_{c}^{\left(\mu\right)}\right\} $
for trees that are not chains (A)} Histogram of the critical values
for trees up to size 15, excluding chains, which are possible subgraphs
of random regular graphs with connectivity $C=3,4,5$ . \textbf{(B)}
Trees can generate jumps in $\phi$ at $\alpha<1/2$: behavior of
$\phi$ for $S=400,C=3$ around the value $\alpha_{c}^{\left(\mu\right)}$
associated with the tree shown.}
\end{figure}

As mentioned in section \ref{subsec:Changes-in-allowed subgraphs},
we find that for a chain of length $n$, 
\begin{equation}
\alpha_{\mathrm{chain}}^{\left(n\right)}=\begin{cases}
\frac{1}{2\cos\left(\frac{\pi}{n+1}\right)} & n\text{ even}\\
\frac{1}{2} & n\text{ odd}
\end{cases}\label{eq: alpha chains appendix}
\end{equation}
and for all other trees, $\alpha_{c}^{\left(\mu\right)}\leq\frac{1}{2}$,
with $\mu$ going over all trees. A histogram of the values for small
trees, calculated numerically, appear in Fig \ref{fig:non-chain histogram}(A),
and the example in Fig. \ref{fig:non-chain histogram}(B) shows that
they do indeed generate jumps in $\phi$ even in the $\alpha<1/2$
region.

While it is fairly easy to determine the critical $\alpha$'s for
chains by checking stability and feasibility directly, it is hard
to show that all trees that are not chains are either unfeasible or
unstable above 1/2. E.g., to show they are unfeasible, we have to
find the solutions to the equilibrium equations and check that one
of the species has a negative abundance, but there is no general formula
for inverting the interaction matrix on an arbitrary tree. Instead
of doing this, we can note that if the tree is unfeasible or unstable,
it will still have a stable equilibrium (because there is a Lyapunov
function so it cannot oscillate indefinitely), but this stable equilibrium
has missing species. It must decompose into subgraphs that are stable
and feasible, which are expected to be only chains above 1/2. Conversely,
the reasoning from sec \ref{subsec:UFP-to-MFP transition} shows that
if such an uninvadable equilibrium made up of chains exists, the full
graph cannot be stable and feasible. This does not require inverting
the matrices for trees, just doing a more combinatorial problem of
splitting the tree into pieces. This idea and another lemma are the
basis for the proof.

These two lemmas, proved in Sec. \ref{subsec:proof Supporting-Lemmas},
are (1) A tree with an unstable sub-tree is itself unstable, and (2)
Any subgraph is feasible and stable if and only if there is no stable
and uninvadable equilibrium on the tree where some of the species
are extinct. We then prove the results on chains, Eq. (\ref{eq: alpha chains appendix})
in Sec. \ref{subsec:proof chains}. Although this can be done directly,
the lemmas give more intuitive derivations of parts of the results.
For example, Lemma 2 allows us to show that an odd length chain is
never feasible and stable above 1/2 by finding an equilibrium with
extinct species in that range. We prove in Sec. \ref{subsec:proof trees ciritcal below half}
that for trees that are not chains $\alpha_{c}^{\left(\mu\right)}\leq1/2$,
in two stages: first, we use lemma 1 to show that unless the junctions
in the tree have at least one neighbor of degree 1 and the rest of
degree 2 at most, the tree is unstable. Next we prove that, for any
junction with these properties, an equilibrium exists for $\alpha>1/2$
in which the species at the junction is extinct, so by lemma 2, the
tree is not feasible and stable.

\subsection{\label{subsec:proof Supporting-Lemmas}Supporting Lemmas}

Here we will introduce two lemmas to aid the proof. They are also
of interest in their own right.

\textbf{Lemma (1): }A graph that has an unstable subgraph is itself
unstable.

Proof: To prove this, denote the interaction matrices of the entire
graph and the subgraph as $\alpha_{ij}$ and $\alpha_{ij}^{*}$ respectively,
and their minimal eigenvalues as $\lambda_{\mathrm{min}},\lambda_{\mathrm{min}}^{*}$,
with $\lambda_{\mathrm{min}}^{*}<0$ as the subgraph is unstable.
As $\alpha_{ij}^{*}$ is a principal submatrix of $\alpha_{ij}$,
from the Cauchy eigenvalue interlacing inequality $\lambda_{\mathrm{min}}\leq\lambda_{\mathrm{min}}^{*}<0$,
meaning the full graph is unstable.

\textbf{Lemma (2): }A graph is feasible and stable if and only if
it there is no stable equilibrium where some of the species are extinct,
that is also uninvadable on the graph.

Proof: If the graph is feasible and stable, then the interaction matrix
for the tree, $\alpha_{ij}$, is positive definite so the Lyapunov
function is concave and the equilibrium is unique; hence there is
no other equilibrium in which some species is extinct. If it is not
feasible and stable, the existence of a Lyapunov function still implies
that there must be some equilibrium, and since the graph is not feasible
and stable some species in this equilibrium must be extinct.

\subsection{\label{subsec:proof chains}Chains}

From Lemma 1 we immediately see that the only allowed subgraph at
$\alpha>1$ is a singlet, as any other subgraph includes a length-2
chain, which is unstable in this range.

In order to calculate the $\alpha_{\text{chain}}^{\left(n\right)}$,
we first derive a rule relating the range of stability and feasibility
to the degrees of the vertices of a graph. Using Lemma 2 we show that
for any subgraph $\mu$, $\alpha_{c}^{\left(\mu\right)}\geq1/\left(\max_{j}C_{j}\right)$,
where $C_{j}$ is the number of interacting neighbors of species $j$.
For a chain, $\max_{j}C_{j}=2$, and so for any $n\ge2$, $\alpha_{\text{chain}}^{\left(n\right)}\geq\frac{1}{2}$.

Indeed, assume that species $j$ is extinct at a fixed point. Its
growth rate is $g_{j}=1-\alpha\sum_{k\sim j}N_{k}\geq1-\alpha C_{j}$.
For $\alpha<1/C_{j}$ the growth rate would be positive and the equilibrium
would be invadable, so in this range species $j$ cannot be extinct
at an equilibrium. So at $\alpha<\min_{j}\left(1/C_{j}\right)$, no
species can be extinct at a fixed point, and at the equilibrium all
species persist. This behavior is apparent in the histogram of values
$\alpha_{c}^{\left(i\right)}$ calculated numerically in Fig. \ref{fig:non-chain histogram}(A)

\textbf{For even length chains, $\alpha_{c}=\left[2\cos\left(\frac{\pi}{n+1}\right)\right]^{-1}$:}
This can be proved by directly checking the feasibility and stability
of the chains above 1/2. The interaction matrix representing a chain
of length $n$ is an $n\times n$ tridiagonal Toeplitz matrix with
1 on the main diagonal and $\alpha$ on the diagonals above and below.
In this case, the $k$th eigenvalue, with $k=1,...,n$, is 
\begin{equation}
\lambda_{k}=1-2\alpha\cos\left(\frac{k\pi}{n+1}\right)\leq\lambda_{1}
\end{equation}
and the chain will be stable for 
\begin{equation}
\alpha<\frac{1}{2\cos\left(\frac{\pi}{n+1}\right)}
\end{equation}
\citep{may_Stability_1973}. The chain is also feasible in this range
if the abundances are all positive. These abundances equal the sum
of the columns of the inverse matrix $\alpha_{ij}^{-1}$. Denoting
its components as $\sigma_{jk}$, for $j\leq k$ they are given by
\begin{equation}
\sigma_{jk}^{n}=\left(-1\right)^{j+k}\frac{1}{\alpha}\frac{U_{j-1}\left(\frac{1}{2\alpha}\right)U_{n-k}\left(\frac{1}{2\alpha}\right)}{U_{n}\left(\frac{1}{2\alpha}\right)}\ ,
\end{equation}
where $U_{m}\left(x\right)$ is the $m$-th Chebyshev polynomials
of the second kind. The sum over the columns reduces to only two entries
of the inverse matrix, so the abundance of the $k$-th species on
a chain of length $n$ is \citep{dafonseca_Explicit_2001} 
\begin{equation}
N_{n,k}=\frac{1+\alpha\left(\sigma_{1k}^{n}+\sigma_{1,n-k+1}^{n}\right)}{1+2\alpha}\ ,
\end{equation}
From Lemma 2 we already obtained that $\alpha_{\text{chain}}^{\left(2m\right)}>\frac{1}{2}$,
and in the region $\alpha>1/2$, the $n$th Chebyshev polynomial would
be $U_{n}\left(\frac{1}{2\alpha}\right)=\frac{\sin\left(\left(n+1\right)\theta\right)}{\sin\theta}$,
where $\cos\theta=\frac{1}{2\alpha}$. In the region where the chain
is stable, $\alpha<1/2\cos\left(\frac{\pi}{n+1}\right)$, it is straightforward
to show that all abundances are positive.

\begin{figure}
\begin{centering}
\includegraphics[width=0.8\columnwidth]{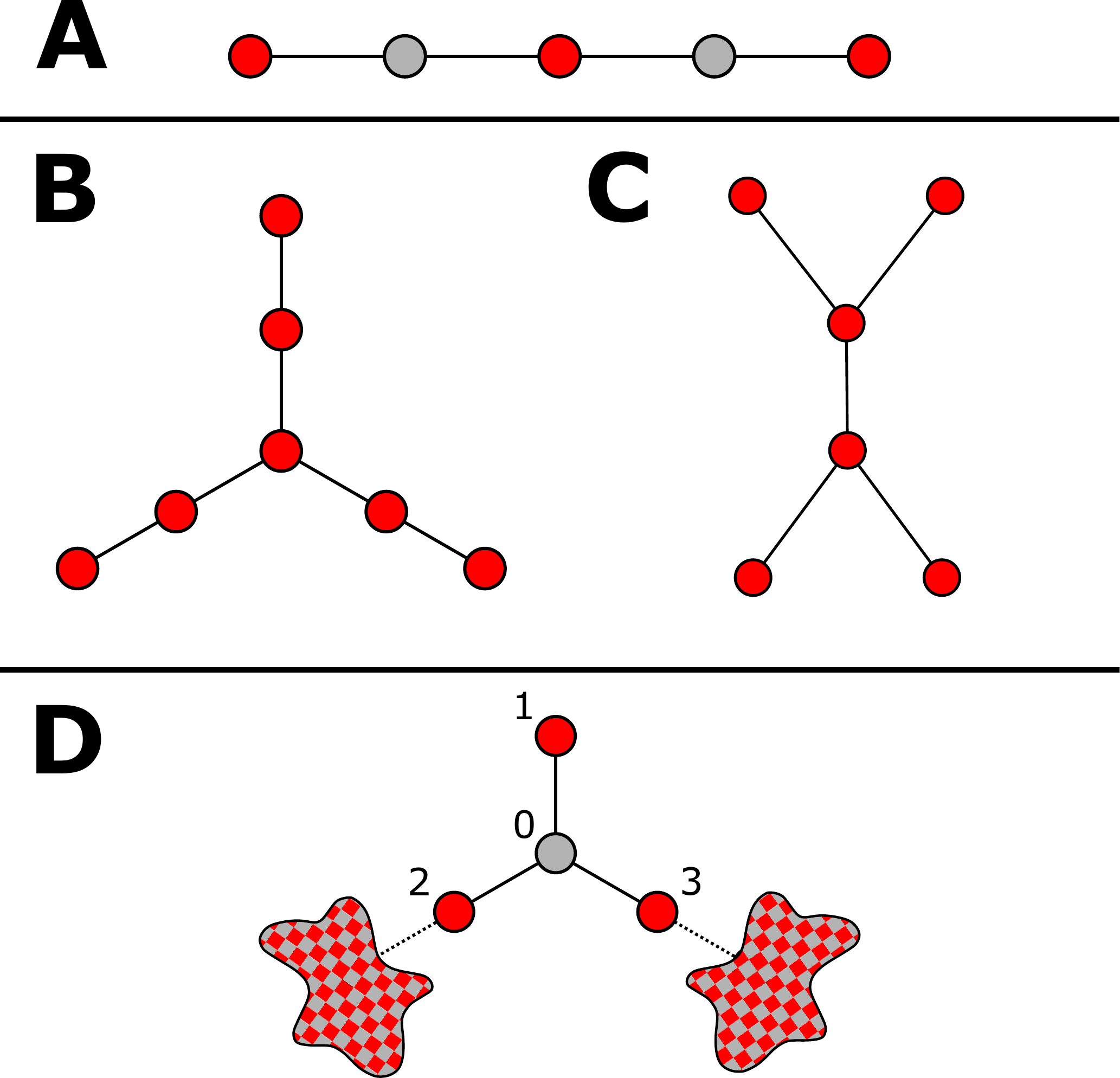}
\par\end{centering}
\caption{\label{fig:tree proof aids}\textbf{(A)} The equilibrium at $\alpha\protect\geq1/2$
for odd length chains includes alternating persistent (red) and extinct
(gray) species, with the persistent species at the odd positions.
\textbf{(B-C)} Trees which become unstable for $\alpha>1/2$. All
trees that have these as subtrees are also unstable at this range.
\textbf{(D)} An equilibrium at $\alpha>1/2$ for all trees that do
not have (B-C) as subtrees. Species 0 as defined in the text, at the
middle of the junction, is extinct. It breaks the tree into several
smaller subtrees of unknown topology, with its neighbors, species
$1,...,m$ in the text, persistent leaves of each subtree. Specifically,
species 1 has no neighbors besides species 0 and is persistent with
$N_{i}=1$.}
\end{figure}

\textbf{For odd length chains, $\alpha_{c}=1/2$:} For $\alpha>1/2$,
a chain of odd length has an uninvadable equilibrium where species
alternate between persistent and extinct, with the persistent species
at the odd positions (See Fig \ref{fig:tree proof aids}(A)). The
persistent species would have only extinct neighbors and therefore
would be stable with abundances $N_{i}=1$. Each extinct species would
have two persistent neighbors and its growth rate will be negative,
$g=1-2\alpha<0$.

\subsection{\label{subsec:proof trees ciritcal below half}Trees that are not
chains have $\alpha_{c}\protect\leq1/2$}

To show that such a tree is not allowed above 1/2, we would like to
use lemma 2 \textendash{} if we can remove vertices such that the
tree breaks up into chains and the removed vertices are kept from
invading by the interactions with the neighbors, we will know that
the tree does not also have a fully populated stable fixed point.
It is natural to try to remove the species at a junction between three
or more branches. We introduce a certain property of junctions and
show that if a junction has this property, then when the species at
the junction is removed, at equilibrium the abundances of its neighbors
are large enough to keep it from invading. If the junction does not
have this property, the tree is unstable and therefore if allowed
to evolve, some species will naturally become extinct.

\textbf{Trees with unstable subtrees:} There are two specific trees,
shown in Fig. \ref{fig:tree proof aids}(B-C), that are unstable exactly
for $\alpha>1/2$, as can be directly verified. The tree in \ref{fig:tree proof aids}(C)
is a subtree of all trees that have two neighboring junctions, and
therefore all such trees have $\alpha_{c}\leq1/2$. Trees that have
no neighboring junctions will still have at least one junction (as
otherwise they would be chains). For such a tree, if all neighbors
of the vertices at the junctions have degree $2$, then the tree \ref{fig:tree proof aids}(B)
is a subtree and it is also unstable at $\alpha\geq1/2$. It remains
to be shown that trees that are not chains and do not contain these
two subtrees also have $\alpha_{c}\geq1/2$. These trees must contain
at least one junction, as they are not chains; vertices neighboring
the junctions can have no more than one other neighbor, otherwise
the tree contains the subtree in Fig. \ref{fig:tree proof aids}(C);
and at least one neighbor of each junction must have no other neighbors,
otherwise the tree contains the subtree in Fig. \ref{fig:tree proof aids}(B).
A visualization of such trees, in the case where the junction has
3 neighbors, is shown in Fig. \ref{fig:tree proof aids}(D).

\textbf{All trees that are not chains have $\alpha_{c}\leq1/2$:}
As we already know that for $\alpha>1$ all trees are unstable, let
us look at trees for some given $1/2<\alpha<1$, and show that they
cannot be stable and feasible at this $\alpha$. We will also use
the fact that for $\alpha<1$, the leaves of a tree are always persistent
at an equilibrium, as a leaf has only a single neighbor, and therefore
if leaf $i$ is extinct its growth rate is $g_{i}=1-\sum_{j}\alpha_{ij}N_{j}\geq1-\alpha>0$.
Further, its abundance is bounded from below, $N_{i}\left(\alpha\right)=1-\alpha\sum_{j\sim i}N_{j}\geq1-\alpha$.

If a tree has either of the trees in Fig \ref{fig:tree proof aids}(B-C)
as subtrees, then we already showed that it is unstable at $\alpha$.
Otherwise, as explained at the end of the previous part, this tree
must have a junction with no neighboring junctions, and at least one
neighbor with degree 1, as shown in Fig \ref{fig:tree proof aids}(D).
We now find an equilibrium with extinct species for trees with such
a topology.

Mark the species in the middle of the junction as species 0, and its
neighbor that has degree 1 as species 1. Species 0 has additional
neighbors $2,...,m$, with $m\geq3$, as indicated in Fig \ref{fig:tree proof aids}(D).
Each of these species has at most one more neighbor besides species
0, otherwise there would be a neighboring junction to species 0.

We now show that an equilibrium exists where species 0 is extinct.
As species 0 goes extinct, the tree separates into $m$ distinct subtrees
of size $<N$. Let us examine the equilibria of these subtrees. The
first subtree includes only species 1 (as it had no other neighbor
besides species 0), so it has an equilibrium where species 1 is persistent
with $N_{1}=1$. All other subtrees also have an equilibrium, because
of the existence of a Lyapunov function for each subtree separately.
We now need only check that species $0$ cannot invade at this equilibrium.

Specifically, as species $2,...,m$ are leaves of their respective
trees, they must be persistent at these equilibria, and have abundances
$N_{i}\left(\alpha\right)\geq1-\alpha$. The growth rate of species
$0$ is therefore
\begin{align*}
g_{0} & =1-\alpha\sum_{i=1}^{m}N_{i}\leq1-\alpha-\alpha\sum_{i=2}^{m}\left(1-\alpha\right)\\
 & \leq1-\alpha\left(1+2\left(1-\alpha\right)\right)\leq0
\end{align*}
and indeed it cannot invade. So the tree has an equilibrium with extinct
species, and thus it cannot be stable and feasible.

\section{\label{sec:Extinct-species}Effects of invadability and non-tree
subgraphs}

In the main text, we explain that the jumps in the relative diversity
$\phi\left(\alpha\right)$ in the region $\alpha>1/2$ result from
changes in the stability and feasibility of trees. We neglect the
effect of changes in the feasibility and stability of subgraphs that
are not trees, and the invadability of extinct species. In this section
we will discuss these assumptions and show that such changes do not
appear to generate additional jumps in $\phi$.

\subsection{Invadability}

In this section, we will show that for $\alpha>1/2$, most extinct
species would not change their invadability within the ranges where
there is no change in feasibility and stability of trees. In the cases
where they do, the jumps generated, if they exist, are so small that
we do not observe them in our simulations

A change in invadability occurs at $\alpha$-values where there is
a sign change of the growth rate of an extinct species, $g_{i}\left(\alpha\right)=1-\alpha\sum_{j\sim i}N_{j}\left(\alpha\right)$.
For $\alpha>1/2$ the only allowed trees are even length chains, with
a finite number of possible abundances $N_{j}\left(\alpha\right)$,
and so a finite number of possible growth rates $g_{i}\left(\alpha\right)$,
depending on the different possible combinations of neighboring species.
For example, in the range where the only allowed trees are singlets
and length 2 chains, the only possible abundances are $N_{i}\left(\alpha\right)\in\left\{ 0,1,\frac{1}{1+\alpha}\right\} $
and for a $C$-regular graph, this gives $\left(\begin{array}{c}
C+2\\
C
\end{array}\right)$ possible growth rates. We checked for such sign changes for $C=3$
in the range that allows chains of up to length 2,4,6 and 8, about
$\alpha\in\left[0.52,1\right]$. We exclude the values $\alpha_{\mathrm{chain}}^{\left(2n\right)}$
where we already know there are jumps in $\phi$ due to changes in
chain stability. We stop at the maximal chain length of 8 as the number
of possible growth rates grows very fast with the maximal allowed
chain length. In the ranges allowing chains of up to length 2 and
4, no growth rate changes sign; in the range allowing chains of up
to length 6, one possible growth rate out of 120 changes sign; in
the range allowing chains of up to length 8, 5 out of 364 possible
growth rates change sign.

In Fig. \ref{fig:invadability and cycles non-jumps}(B), we show $\phi\left(\alpha\right)$
around a value of $\alpha$ where one of these changes in growth rates
occur, along with the specific combination of neighbors that generates
the change. Even with an increase in the pool size $S$, we see no
jump in the value of $\phi$ within statistical error. These results
could be expected, as each growth rate occurs only when an extinct
species has neighboring chains of very specific lengths , which happens
very infrequently. However, this does not mean that invadability is
unimportant, as it drives the changes in the allowed subgraphs. For
example, chains of length 4 become allowed at the value of $\alpha$
such that an extinct species which neighbors a length two chain (of
persistent species) and another single species can invade, so that
all the sites stick together as a length 4 chain. This is just another
way of describing the result above, that a graph is not allowed if
removing some species from it leads to a subgraph such that the removed
species cannot invade.

\subsection{Non-tree subgraphs}

\begin{figure}
\begin{centering}
\includegraphics[width=0.8\columnwidth]{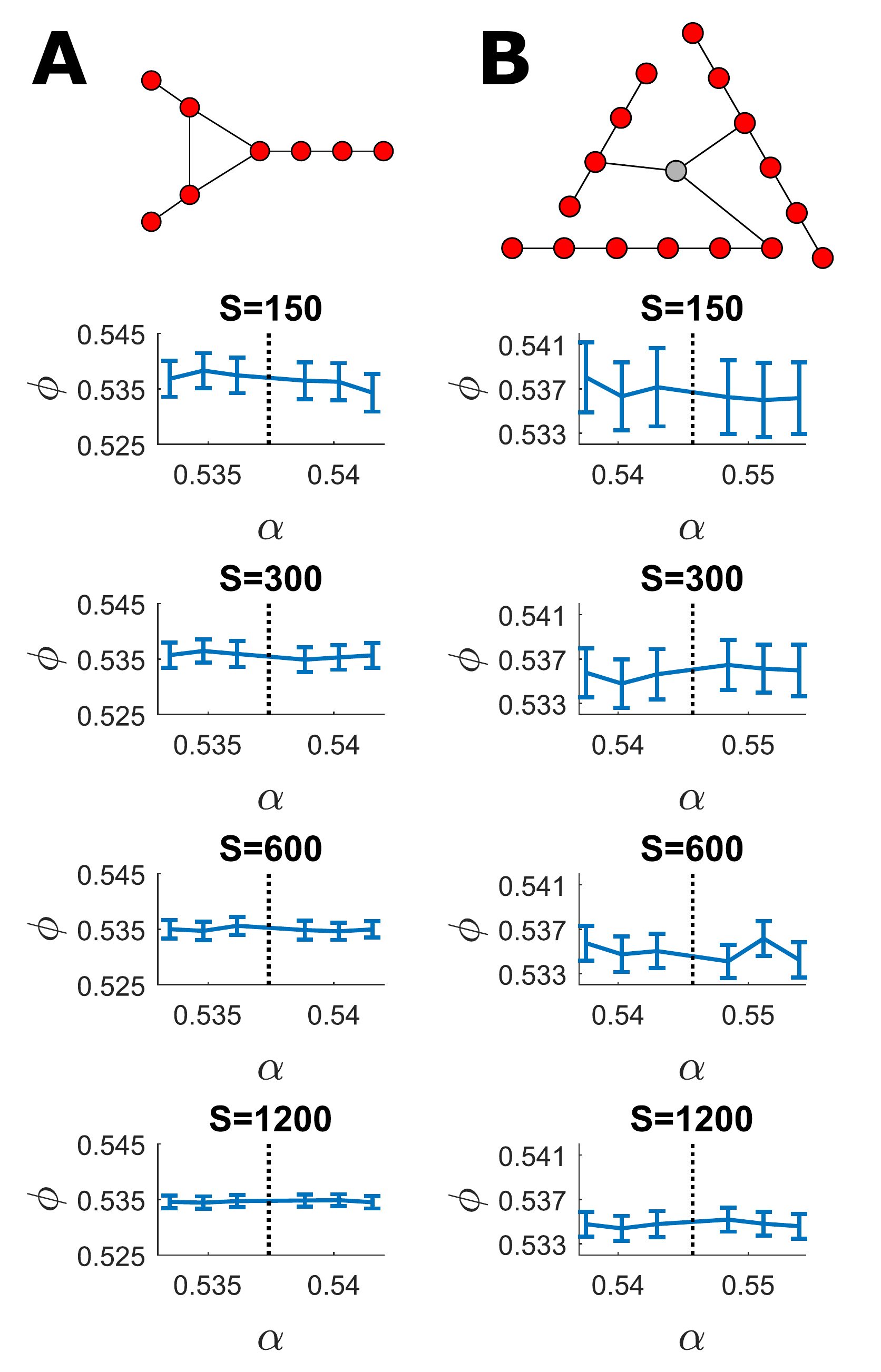}
\par\end{centering}
\caption{\label{fig:invadability and cycles non-jumps}\textbf{Non-tree subgraphs
and invadability changes do not generate jumps in $\phi$ at $\alpha>1/2$.}
The dependence of relative diversity $\phi$ on the interaction strength
$\alpha$ for increasing pool sizes $S$, in two cases \textbf{(A)}
Around $\alpha\approx0.537$, the value at which a non-tree subgraph
becomes allowed. The subgraph is shown at the top \textbf{(B)} Around
$\alpha\approx0.546$, the value where an extinct species (in gray)
with a set of persistent neighbors as shown will change the sign of
its growth rate, and so its invadability.}
\end{figure}

As mentioned in the main text, as sparse graphs are tree-like and
short cycles are rare, we expect to see no jumps generated by subgraphs
that are not trees. Fig. \ref{fig:invadability and cycles non-jumps}(A)
shows an example for a specific subgraph that includes a cycle: $\phi\left(\alpha\right)$
displays no jump around the critical value where this subgraph becomes
allowed, even as we increase $S$.

\section{\label{sec:C Global transitions with heterogeneity}Collective transitions
with heterogeneity}

\begin{figure}
\begin{centering}
\includegraphics[width=1\columnwidth]{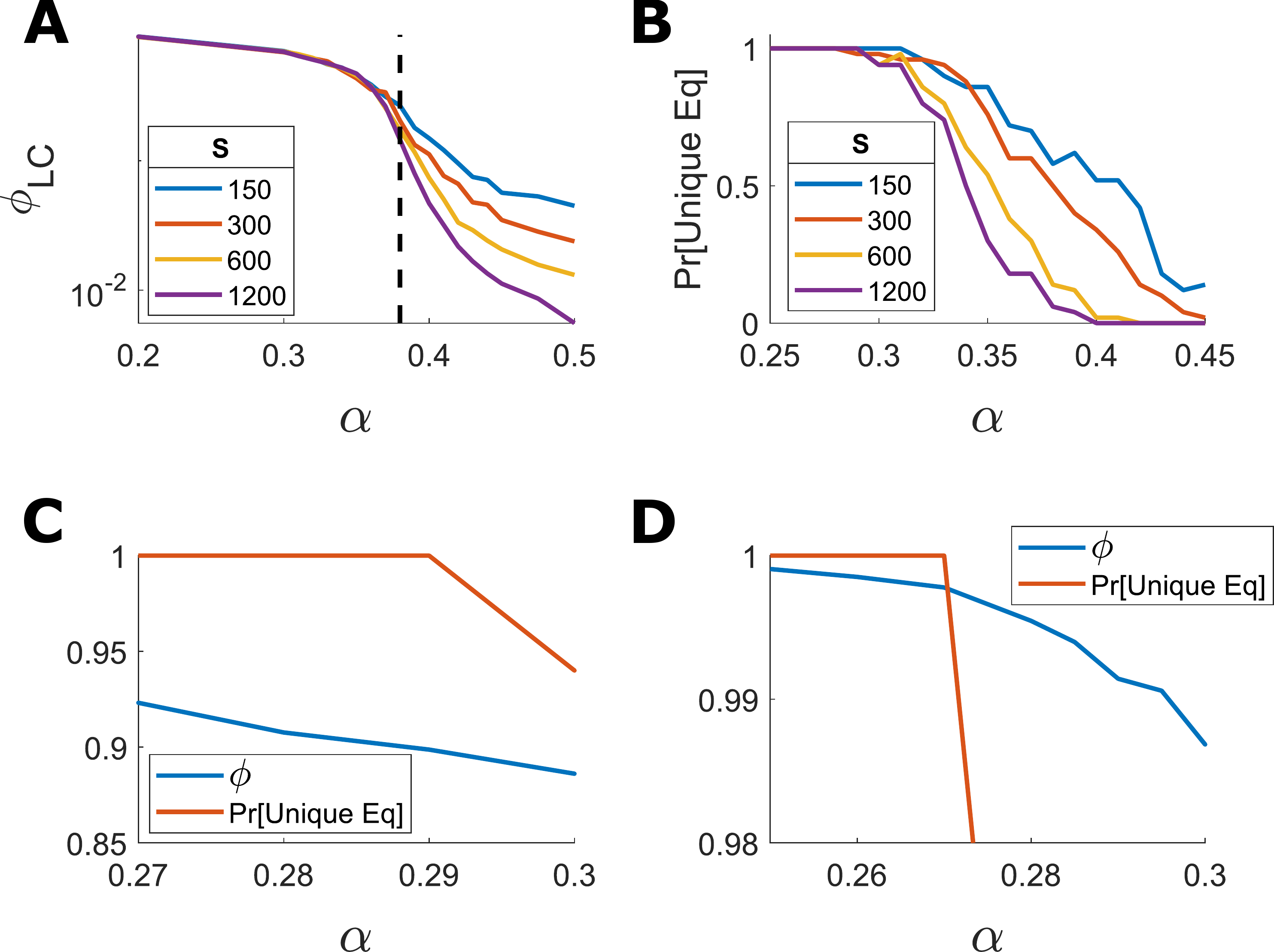}
\par\end{centering}
\caption{\label{fig:percolation and UFP/MFP with noise}\textbf{Collective
transitions with heterogeneity.} Results in panels (A-B) are the equivalent
of Fig. \ref{fig:UFP-MFP and PERC phase transition}(C,D), here for
heterogeneity in degree the transitions. As in Fig. \ref{fig:UFP-MFP and PERC phase transition}(C,D),
the transitions become sharper as $S$ increases, within the range
of $S$ tested numerically. They show simulation results for Erd\H{o}s-Rényi
graphs with average degree\textbf{ $C=3$} and several values of $S$.
\textbf{(A) }Percolation transition: The fraction of species in the
largest connected component as a function of $\alpha$. At low values
of $\alpha$ a finite fraction of species belong to the largest component,
and above a transition the fraction of species decreases with $S$.
\textbf{(B)} Multiple to unique equilibrium transition: The probability
of having a unique equilibrium as a function of $\alpha$. \textbf{(C-D)}
For both types of heterogeneity some species go extinct (a loss of
feasibility of the entire system) before loss of stability: The fraction
of surviving species $\phi$ drops below 1 in the unique equilibrium
phase. Results are shown for $S=1200,C=3$. The probability for a
unique equilibrium is shown in red, and $\phi$ in blue. \textbf{(C)}
Heterogeneity in degree, Erd\H{o}s-Rényi graphs \textbf{(D) }Heterogeneity
in interaction strength, $\sigma=0.1$.}
\end{figure}
In this section we continue to examine the two collective transitions,
the transition from multiple to unique equilibria and the percolation
transition, in cases where interaction strengths and vertex degrees
are not constant across the network. The behavior at the transitions
is shown in Fig. (\ref{fig:phi of alpha with noise}) in the main
text for heterogeneous interaction strengths, and here in the top
panels of Fig (\ref{fig:percolation and UFP/MFP with noise}) for
variability in vertex degree modeled by an Erd\H{o}s-Rényi graph.
For both cases, the transition from multiple to unique equilibria
becomes sharper as $S$ increases (within the range checked numerically),
with the probability of a unique equilibrium approaching a step function.
The percolation transition in both cases is qualitatively similar
to the transition that occurs in the case with no heterogeneity, as
well as to standard site percolation, see Section \ref{sec:classical percolcation}.

Fig. \ref{fig:percolation and UFP/MFP with noise}(C,D) show that
for both types of heterogeneity, $\phi$ drops below 1 before the
transition to multiple equilibria, for $\alpha<\alpha_{\text{UE}}$.
Therefore, the feasibility of the entire system is lost before its
stability. For heterogeneous interaction strengths, this follows from
Lemma (3) in Appendix \ref{sec:App-Hierarchy}.

\section{\label{sec:classical percolcation}Comparison to standard percolation}

\begin{figure}
\centering{}\includegraphics[width=1\columnwidth]{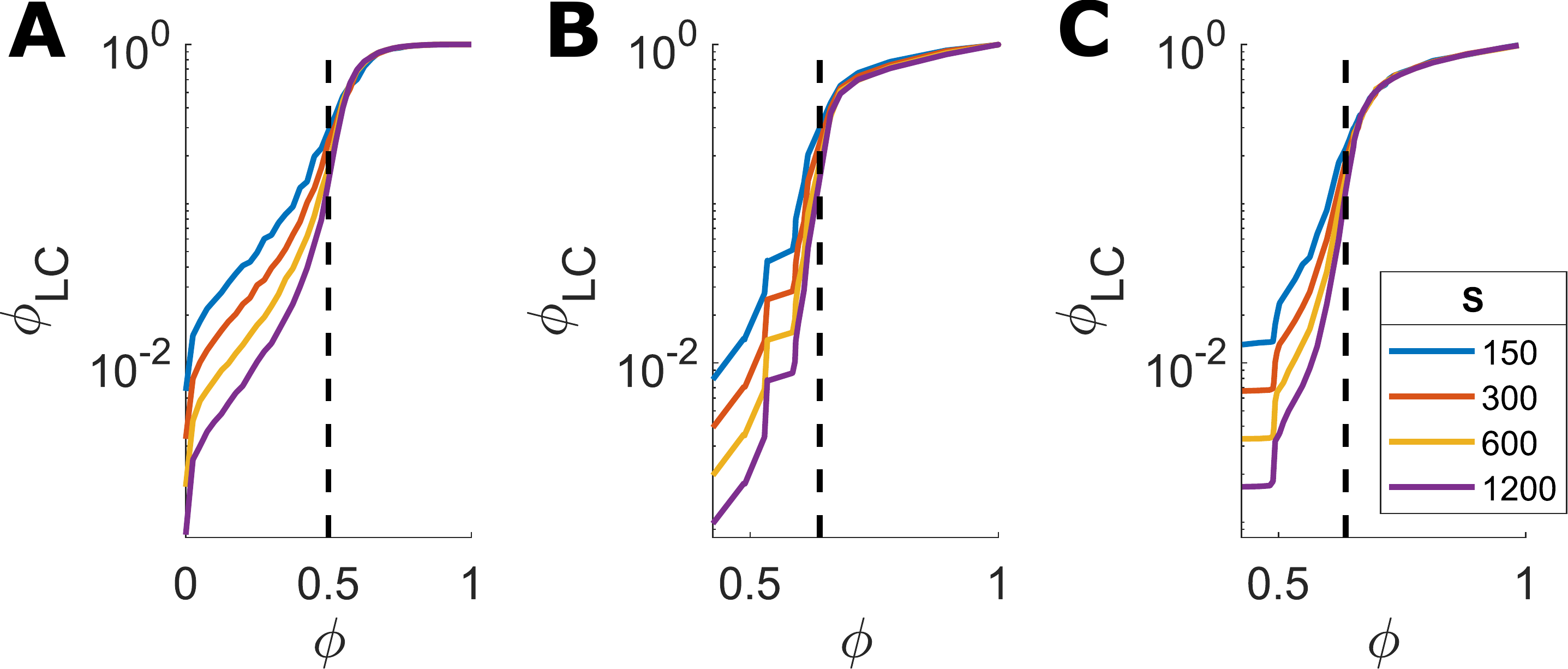}\caption{\label{fig:classical percolation}\textbf{Percolation in our model
is qualitatively similar to standard site-percolation, near the transition.
}The dependence of the fraction of species in the largest connected
component on the fraction of all persistent species $\phi$, for $C=3$
and several pool sizes $S$. \textbf{(A) }Standard site-percolation:
vertices are taken to be present with probability $\phi$, independently
for each vertex. The percolation transition occurs at $\phi_{\text{perc}}=1/2$,
where $\phi_{LC}\sim S^{-1/3}$. \textbf{(B)} The equal-$\alpha$
model, with constant interaction strength and vertex degree, $\phi_{\text{perc}}\approx0.64$.
\textbf{(C)} The model with variability $\sigma=0.1$ in interaction
strength, $\phi_{\text{perc}}\approx0.636$.}
\end{figure}

Here we elaborate on the comparison in the main text between percolation
in our model and standard site percolation. In standard percolation,
each vertex is taken to be ``present'' with a given probability
$p$, and for $C$-regular graphs the percolation transition is known
to occur at $p_{\text{perc}}\left(C\right)=\frac{1}{C-1}$ \citep{bunde_Fractals_1996}.
Fig. \ref{fig:classical percolation} compares three cases: standard
percolation on a random regular graph, percolation in the equal-$\alpha$
model where interaction strength and degree are constant, and for
heterogeneous interaction strengths. For each we show the dependence
of the fraction of species in the largest connected component, $\phi_{LC}$,
on the fraction of surviving species, $\phi$, or on $p$ for standard
percolation. As mentioned in the main text, in all cases the behavior
close to the transition is qualitatively similar. We use this similarity
to estimate $\alpha_{\text{perc}}$ in our model, as the value of
$\alpha$ where the fraction of species in the largest component grows
as $S^{-1/3}$, as in known to occur at $p_{\text{perc}}$ for standard
site percolation \citep{bunde_Fractals_1996}. As mentioned, for our
model $\phi_{\text{perc}}>1/2=p_{\text{perc}}$, due to the fact that
persistent species are anticorrelated, tending not to be adjacent
to one another.

\section{Subgraph emergence rule\label{sec:App-Hierarchy}}

To prove the result quoted in the main text, we first prove a Lemma,
which is interesting in its own right. We use the term ``generically''
for ``with probability approaching one for large numbers of species''.

\textbf{Lemma (3)}: Consider a system with symmetric ($\alpha_{ij}=\alpha_{ji}$)
and competitive ($\alpha_{ij}\ge0$) interactions, sampled from some
continuous distribution (such as a Gaussian distribution as in the
main text). Suppose that the $\alpha_{ij}$ are changed continuously
by shifting $m\equiv\text{mean}\left(\alpha_{ij}\right)$ (other continuous
shifts are also possible). Assume that the graph is feasible and stable
in some range below $m=\alpha_{c}$, and not in a range above it.
Then generically, it is feasibility that breaks at $\alpha_{c}$,
by a single species' abundance going to zero, while stability continues
to hold.

Proof: We prove this by contradiction. Assume to the contrary that
at $m=\alpha_{c}$ the graph becomes unstable. As the matrix $\alpha$
is symmetric it can be diagonalized. Let $\alpha=\sum_{i}\lambda_{i}\vec{v}_{i}\vec{v}_{i}^{T}$
be its eigen-decomposition, where $T$ denotes the transpose operation,
$\left\{ \vec{v}_{i}\right\} $ are column eigenvectors, and $\lambda_{i}$
the corresponding eigenvalues, with $\lambda_{1}<\lambda_{2}<...$
(generically there is no degeneracy). Note that the values of quantities
in this decomposition depend on $m$. These values are equilibrium
solutions to Eq. (\ref{eq:LV}); fromfeasibility up to $\alpha_{c}$,
all $N_{i}>0$ so,
\[
\vec{N}=\alpha^{-1}\vec{u}=\Sigma_{j}\lambda_{i}^{-1}\vec{v}_{i}\vec{v}_{i}^{T}\vec{u}\ ,
\]
where $\vec{u}=\left(1,1,..\right)$. By assumption, the system becomes
unstable at $\alpha_{c}$, $\lambda_{1}\stackrel{m\rightarrow\alpha_{c}}{\longrightarrow}0$.
Since generically $\vec{v}_{i}^{T}\vec{u}\ne0$, the first term dominates
near $\alpha_{c}$,
\[
\vec{N}=\lambda_{1}^{-1}\vec{v}_{1}\left(\vec{v}_{1}^{T}\vec{u}\right)+\left\langle \mathrm{terms\ finite\ as\ \alpha\rightarrow\alpha_{c}}\right\rangle \ ,
\]
so $\vec{N}$ diverges at $m\rightarrow\alpha_{c}$. Using $\alpha_{ij}\ge0$
and feasibility, $N_{i}=1-\sum_{j}\alpha_{ij}N_{j}\le1$. Therefore,
the divergence of the values of $\vec{N}$ must be towards $-\infty$,
and so the $N_{i}$-values must cross zero at $m$ \emph{smaller than}
$\alpha_{c}$, in contradiction to the assumption. QED

Applying this lemma, a subgraph that loses feasibility at $\alpha_{c}$
generically does so by only one species having $N_{i}\rightarrow0$.
The remaining graph is still feasible and stable at $\alpha_{c}$
and for at least some range $\left[\alpha_{c},\alpha_{c}+\varepsilon\right]$
above it (because the stability and abundances of the remaining species
change continuously). In the case of trees, removing a vertex splits
the tree into multiple trees, see Fig. \ref{fig:Hierarchy}.

Without heterogeneity (when all $\alpha_{ij}=\alpha$) all trees have
$\alpha_{c}\leq1/2$, so it is interesting to consider the case where
all the $\alpha_{ij}$ connecting to the extinct species $N_{i}$
satisfy $\alpha_{ij}<1/2$. In this case, the extinct species has
$0=N_{i}=1-\sum_{j}\alpha_{ij}N_{j}>1-C/2$, where $C$ is the degree
of species $i$, so $C>2$ and the tree will split into at least three
parts.

\bibliographystyle{unsrt}
\bibliography{Symmetric_interactions_paper}

\end{document}